\documentstyle[twoside,leqno]{article}

\newif\ifdopreprint		\dopreprinttrue
\newif\ifdopsdraft		\dopsdraftfalse
\newif\ifdodoublespace		\dodoublespacetrue

\ifdopreprint
\makeatletter
\def\l@section#1#2{\addpenalty{\@secpenalty}
\@tempdima 1.5em \begingroup
 \parindent \z@ \rightskip \@pnumwidth
 \parfillskip -\@pnumwidth
 \bf \leavevmode \advance\leftskip\@tempdima \hskip -\leftskip #1\nobreak\hfil
\nobreak\hbox to\@pnumwidth{\hss #2}\par
 \endgroup}
\makeatother
\fi

\input{psfig}

\ifdopsdraft
\psdraft     
\fi

\newtheorem{theorem}{Theorem}[section]
\newtheorem{lemma}[theorem]{Lemma}

\newtheorem{definition}[theorem]{Definition}

\newenvironment{proof}{\medskip \noindent {\bf Proof:}}%
                      {$\quad \Box$ \smallskip}

\def\Kronecker{{Kr\"{o}necker }}

\def\R{{\bf R}}
\def\RD{{\R^D}}
\def\E{{\bf E}}
\def\P{{P}}

\def\X{{\bf X}}
\def\L{{\bf L}}
\def\Two{{\bf 2}}
\def\TwoP{{\Two^\P}}

\def\eps{{\epsilon}}

\def\x{{\bf x}}
\def\v{{\bf v}}
\def\i{{\bf i}}
\def\q{{\bf q}}
\def\p{{\bf p}}

\def\AA{{\cal A}}
\def\BB{{\cal B}}
\def\CC{{\cal C}}
\def\DD{{\cal D}}
\def\EE{{\cal E}}
\def\II{{\cal I}}
\def\JJ{{\cal J}}
\def\LL{{\cal L}}
\def\MM{{\bf M}}
\def\PP{{\widehat{P}}}  
\def\SS{{\cal S}}
\def\TT{{\cal T}}

\def\KA{C}		

\def\evec{{\vec{e}}}
\def\betavec{{\vec{\beta}}}
\def\alphavec{{\vec{\alpha}}}

\def\FBar{{\bar{F}}}
\def\MBar{{\bar{M}}}
\def\NBar{{\bar{N}}}
\def\nBar{{\bar{n}}}

\def \M  {{\!_M}}

\def\del   {{\partial}}
\def\GRAD  {{\nabla}}
\def\DIV   {{\nabla\cdot}}

\def\NULL  {\mathop{\rm Null}}

\def\SPAN  {\mathop{\rm Span}}

\def\<{{\langle}}
\def\>{{\rangle}}

\def\frac#1#2{{#1 \over #2}}
\def\tfrac#1#2{{\textstyle{#1 \over #2}}}

\makeatletter
\newcount\@minsofar
\newcount\@min
\newcount\@cite@temp
\def\@citex[#1]#2{%
\if@filesw \immediate \write \@auxout {\string \citation {#2}}\fi
\@tempcntb\m@ne \let\@h@ld\relax \def\@citea{}%
\@min\m@ne%
\@cite{%
  \@for \@citeb:=#2\do {\@ifundefined {b@\@citeb}%
    {\@h@ld\@citea\@tempcntb\m@ne{\bf ?}%
    \@warning {Citation `\@citeb ' on page \thepage \space undefined}}%
{\@minsofar\z@ \@for \@scan@cites:=#2\do {%
  \@ifundefined{b@\@scan@cites}%
    {\@cite@temp\m@ne}
    {\@cite@temp\number\csname b@\@scan@cites \endcsname \relax}%
\ifnum\@cite@temp > \@min
    \ifnum\@minsofar = \z@
      \@minsofar\number\@cite@temp
      \edef\@scan@copy{\@scan@cites}\else
    \ifnum\@cite@temp < \@minsofar
      \@minsofar\number\@cite@temp
      \edef\@scan@copy{\@scan@cites}\fi\fi\fi}\@tempcnta\@min
  \ifnum\@minsofar > \z@ 
    \advance\@tempcnta\@ne
    \@min\@minsofar
    \ifnum\@tempcnta=\@minsofar 
      \ifx\@h@ld\relax
        \edef \@h@ld{\@citea\csname b@\@scan@copy\endcsname}%
      \else \edef\@h@ld{\ifmmode{-}\else--\fi\csname b@\@scan@copy\endcsname}%
      \fi
    \else \@h@ld\@citea\csname b@\@scan@copy\endcsname
          \let\@h@ld\relax
  \fi 
\fi}%
\def\@citea{,\penalty\@highpenalty\,}}\@h@ld}{#1}}

\makeatother

\makeatletter
\def\@abssec#1{\vspace{.05in}\footnotesize \parindent .2in
{\bf #1. }\ignorespaces}
\def\abstract{\@abssec{Abstract}}
\def\keywords{\@abssec{Key words}}
\def\AMSMOS{\@abssec{AMS(MOS) subject classifications}}

\makeatother

\makeatletter

\def\baselinestretch{2}

\def\setstretch#1{\renewcommand{\baselinestretch}{#1}}

\def\singlespace{\def\baselinestretch{1}\@normalsize}

\def\@setsize#1#2#3#4{\@nomath#1\let\@currsize#1\baselineskip
   #2\baselineskip\baselinestretch\baselineskip
   \setbox\strutbox\hbox{\vrule height.7\baselineskip
      depth.3\baselineskip width\z@}
   \normalbaselineskip\baselineskip#3#4}

\skip\footins 20pt plus4pt minus4pt

\def\@xfloat#1[#2]{\ifhmode \@bsphack\@floatpenalty -\@Mii\else
   \@floatpenalty-\@Miii\fi\def\@captype{#1}\ifinner
      \@parmoderr\@floatpenalty\z@
    \else\@next\@currbox\@freelist{\@tempcnta\csname ftype@#1\endcsname
       \multiply\@tempcnta\@xxxii\advance\@tempcnta\sixt@@n
       \@tfor \@tempa :=#2\do
                        {\if\@tempa h\advance\@tempcnta \@ne\fi
                         \if\@tempa t\advance\@tempcnta \tw@\fi
                         \if\@tempa b\advance\@tempcnta 4\relax\fi
                         \if\@tempa p\advance\@tempcnta 8\relax\fi
         }\global\count\@currbox\@tempcnta}\@fltovf\fi
    \global\setbox\@currbox\vbox\bgroup
    \def\baselinestretch{1}\@normalsize
    \boxmaxdepth\z@
    \hsize\columnwidth \@parboxrestore}
\long\def\@footnotetext#1{\insert\footins{\def\baselinestretch{1}\footnotesize
    \interlinepenalty\interfootnotelinepenalty
    \splittopskip\footnotesep
    \splitmaxdepth \dp\strutbox \floatingpenalty \@MM
    \hsize\columnwidth \@parboxrestore
   \edef\@currentlabel{\csname p@footnote\endcsname\@thefnmark}\@makefntext
    {\rule{\z@}{\footnotesep}\ignorespaces
      #1\strut}}}

\makeatother

\begin{document}


\begin{titlepage}\samepage
\setstretch{1}\large\normalsize\rm

\title{%
	\ifdopreprint\vspace{-0.5in}\fi%
	Convergence of Convective-Diffusive Lattice Boltzmann Methods%
}

\author{Bracy H. Elton
        \thanks{ This work was supported in part under the auspices of the
                 U.S.\ Dept.\ of Energy by Lawrence Livermore National
                 Laboratory under contract No.\ W--7405--Eng--48.} \\
        Computational Research Division  \\
        Fujitsu America, Inc. \\
        3055 Orchard Drive  \\
        San Jose, CA~~95134--2022
\and
        C. David Levermore
        \thanks{ This work was performed under the support of NSF
                 grant No.\ DMS--8914420.} \\
        Department of Mathematics  \\
        University of Arizona  \\
        Tucson, AZ~~85721
\and
        Garry H. Rodrigue
        \thanks{ This work was supported under the auspices of the
                 U.S.\ Dept.\ of Energy by Lawrence Livermore National
                 Laboratory under contract No.\ W--7405--Eng--48.} \\
        Department of Applied Science \\
        University of California at Davis and \\
        Lawrence Livermore National Laboratory  \\
        P.O.\ Box 808, L--540  \\
        Livermore, CA~~94551}

\date{March 18, 1993}

\maketitle

\ifdopreprint
\renewcommand{\thepage}{}
\thispagestyle{myheadings}
\def\ReportNumber{\protect\parbox[b]{5in}{%
  \protect\raggedleft\protect\vspace{-\baselineskip}%
     Lawrence Livermore National Laboratory \#UCRL--JC--113509 \protect\\
     Submitted to the SIAM Journal on Numerical Analysis%
  }
}
\markboth{\ReportNumber}{\ReportNumber}
\else
\thispagestyle{empty}
\fi

\ifdopreprint
\setstretch{1}\large\normalsize\rm
\else
\ifdodoublespace
\setstretch{1.37}\large\normalsize\rm
\else
\fi
\fi

\begin{keywords}
consistency,
convection-diffusion,
convergence,
explicit,
Hilbert expansion,
finite difference,
lattice Boltzmann,
lattice gas,
monotone,
stability
\end{keywords}

\begin{AMSMOS}
35F25, 65M06, 65M12, 76R99, 82C40
\end{AMSMOS}

\begin{abstract}
  Lattice Boltzmann methods are numerical schemes derived as a kinetic
approximation of an underlying lattice gas.  A numerical convergence
theory for nonlinear convective-diffusive lattice Boltzmann methods
is established.  Convergence, consistency, and stability are defined
through truncated Hilbert expansions.  In this setting it is shown
that consistency and stability imply convergence.  Monotone lattice
Boltzmann methods are defined and shown to be stable, hence convergent
when consistent.  Examples of diffusive and convective-diffusive
lattice Boltzmann methods that are both consistent and monotone are
presented.

\end{abstract}

\ifdopreprint
{\vfill\samepage\small\rm\tableofcontents\vfill}
\fi

\end{titlepage}

\ifdopreprint
\else
	\clearpage
	\thispagestyle{empty}
	\setstretch{1}\large\normalsize\rm
	\tableofcontents
\fi

\clearpage

\setcounter{page}{1}

\pagestyle{myheadings}

\markboth
{\protect\small\protect\rm
	\hfill{B.H.\ ELTON, C.D.\ LEVERMORE, G.H.\ RODRIGUE}\hfill
}
{\protect\small\protect\rm
	\hfill{CONVERGENCE OF CONVECTIVE-DIFFUSIVE LATTICE BOLTZMANN METHODS}\hfill
}


\ifdodoublespace
\setstretch{1.37}\large\normalsize\rm
\else
\setstretch{1}\large\normalsize\rm
\fi

\section{Introduction}

  Lattice gases, which were introduced in the early 1970s
\cite{HP72,HPP73}, have been used to simulate problems in fluid
dynamics \cite{Doolen91,FHP86,FHHLPR87}.  A Lattice gas involves
indistinguishable pseudo-particles that traverse from node to node along
the links of a lattice in unison according to the ticks of a discrete
clock and that interact at the nodes of the lattice.  An exclusion
principle is imposed so that the state at any given node may be
described with a finite number of bits.  Thus, lattice gases are
amenable to a mathematical description over a Boolean field and have
been related to cellular automata
\cite{von.Neumann!cellular.automata,Wolfram86}.  The microscopic
evolution of a lattice gas system can be viewed as a space-time-velocity
discretization of the Boltzmann equation (see, e.g., \cite{BPT88}), in
which the precision of the particle distributions is reduced to as few
as one bit.  Characteristic of lattice gas methods is that the velocity
discretization remains fixed (to the lattice structure) in the limit as
the spatial and temporal discretization parameters tend toward zero.
While the microdynamics of a lattice gas is certainly not physical,
the aim is however to recover physical macrodynamics via this
simple, non-physical microdynamic means.  This and the accompanying
statistics have been explored for a variety of lattice gas methods
\cite{Doolen91}.  In lattice Boltzmann methods particle distributions
(not particles) traverse the links of the lattice and interact at the
nodes thereof \cite{MZ88,Muir87,SBH89} (see, e.g., \cite{Doolen91}, for
further references).  Lattice Boltzmann methods do not possess the
statistical fluctuations that are inherent in lattice gas methods. For
certain classes of these methods we develop a convergence theory.
Such classes include linear and nonlinear convective-diffusive
and monotone lattice Boltzmann methods.

  Our paper is organized as follows.  In the next section we quantify
the microscopic dynamics of lattice gases and then derive formally
an equation for the expected or mean behavior of this system, the
so-called lattice Boltzmann equation, which forms the basis of lattice
Boltzmann methods.  This equation has the form of a discrete space-time
kinetic equation composed of an advection part and a collision part, the
so-called Boltzmann collision operator, whose properties are examined in
section 3.  The main thrust of this paper is to establish a convergence
theory for solutions of this equation in the continuum limit for a class
of convective-diffusive lattice Boltzmann methods.  In section 4, we
identify the class of lattice Boltzmann methods that have a
convective-diffusive continuum limit through an analogue of the
classical Hilbert expansion of kinetic theory.  This is the lattice
Boltzmann equivalent of the consistency step of traditional convergence
proofs for numerical schemes.  Stability, and therefore convergence, is
then established in section 5 for a class of so-called monotone lattice
Boltzmann methods.  Specific examples of both diffusive and
convective-diffusive lattice Boltzmann methods that are both consistent
and monotone are presented in sections 6 and 7.


\section{Lattice Gas Dynamics}

  A lattice gas \cite{Doolen91} involves indistinguishable particles
moving about from node to node on a lattice in unison with the ticks of
a discrete clock and interacting at the nodes of the lattice.  More
precisely, each particle is characterized as being in one of a finite
number of possible particle states, and associated with each possible
state is a velocity, which is the lattice vector on which the particle
will translate during the advection step of each clock cycle.  Before
the advection step of each cycle, the gas undergoes a collision step
during which the particles at each node interact according to a set of
rules that change the states of the particles at that node independently
of the state of the particles at any of the other nodes.  These
collision rules may be either deterministic or stochastic.

  A lattice gas automaton \cite{Wolfram86} envokes the exclusion
principle, which states that at any given time there is at most one
particle per particle state per node.  This principle ensures that a
single bit, called an occupation number, can encode the absence (=0) or
presence (=1) of a particle in a particular particle state at a given
node.  Thus, a finite number of bits may be used to describe the local
state of the gas at a given node --- one bit for each possible particle
state.

  More specifically, a lattice gas automaton and its dynamics are
quantified as follows \cite{FHP86}:

\begin{enumerate}

\item  A spatial {\bf lattice domain}, $\X\subset\R^D$.  More
       precisely, given a macroscopic domain $\Omega\subset\R^D$,
       set $\X=\Omega\cap\L$ for some regular $D$-dimensional lattice
       $\L\subset\R^D$ with microscopic spacing $\Delta x$.  The nodes
       of $\X$ are denoted $\i$.  In order to avoid the complications
       wrought by boundaries, we shall assume that $\X$ and $\Omega$
       are effectively without boundary by imposing periodicity, say
       of length $L$.

\item  The ticks of the discrete clock are called {\bf cycles} and
       are indexed by $m=0,1,2,\ldots\,$.  Each cycle corresponds to a
       microscopic time step of $\Delta t$.

\item  A finite set $\P$ (of cardinality $|\P|$) of possible
       {\bf particle states} at each node.  A mapping $\v:\P\to\L$
       associates a velocity vector $\v(p)$ with values in a lattice
       neighborhood of the origin to each particle state $p\in\P$.
       In the absence of collisions, a particle in state $p$ will
       translate each cycle by $\v(p)$ along the lattice.

\item  The absence (=0) or presence (=1) of a particle in a particular
       particle state $p$ at the node $\i$ after cycle $m$ is encoded
       by an {\bf occupation number} $N(m,\i,p)\in\Two\equiv\{0,1\}$.
       The local state at the node $\i$ after cycle $m$ is given by
       $N(m,\i,\cdot)\in\TwoP$.  (A ``$\cdot$'' in an argument denotes
       a function over the dotted argument.)

\item  The {\bf advection operator} $\AA$ translates the particles to
       neighboring nodes and advances the discrete time cycle from $m$
       to $m+1$.  It is defined by
$$
  \AA N(m,\i,p) \equiv N(m+1,\i+\v(p),p) \,.
$$

\item  The {\bf collision operator} $\widetilde{\CC}$ acts locally
       in space-time lattice to determine the change in local state
       due to interactions between particles.  More specifically,
       given the local state $N(m,\i,\cdot)$ at node $\i$ after cycle
       $m$, the collision operator determines a
       collided state $N'(m,\i,\cdot)$ by
$$
  N'(m,\i,\cdot) \equiv N(m,\i,\cdot)
                        + \widetilde{\CC}(N(m,\i,\cdot)) \,.
$$
       In other words, the map $n\mapsto n'=n+\widetilde{\CC}(n)$ takes
       the set $\TwoP$ of local states into itself.

\item  The composition of the advection step with the collision step
       gives the {\bf microdynamical equation} for cycle $m+1$ as
$$
  \AA N(m,\i,p) = N(m,\i,p) + \widetilde{\CC}(N(m,\i,\cdot))(p) \,,
$$
       or, more simply,
\begin{equation}
  \label{lgm}
  \AA N = N + \widetilde{\CC}(N) \,.
\end{equation}
       This equation states that after cycle $m+1$ the new occupation
       number for state $p$ at the new location $\i+\v(p)$ is the same
       as the occupation number for state $p$ at the location $\i$
       after cycle $m$, plus some collisional correction.

\end{enumerate}

  Since $N(m,\i,\cdot) \in \TwoP$, it can take on one of $2^{|\P|}$
possible local states in $\TwoP$.  In order to detect exactly which
state is occupied, the general expression for the collision operator
requires a representation of the \Kronecker delta.  Let $N,n\in\TwoP$
have components $N(p),n(p)$ and define
$$
  N^n = \prod_{p \in \P} N(p)^{n(p)} \,, \qquad
  \NBar^\nBar = \prod_{p \in \P} (1-N(p))^{(1-n(p))} \,.
$$
Notice that $N^n\NBar^\nBar=\delta(N,n)$, the \Kronecker delta.
The collision rules of the lattice gas determine a unique
post-collisional state $n'$ for any given precollisional state
$n\in\TwoP$.  Introduce a matrix $\widetilde{\SS}(n,n')$ such that
$$
  \widetilde{\SS}(n,n')
  = \left\{ \begin{array}{ll}
            1 \,, & \hbox{ if $n \mapsto n'$} \,; \\
            0 \,, & \hbox{ otherwise} \,.
            \end{array} \right.
$$
The corresponding collision operator can then be expressed as
$$
  \widetilde{\CC}(N)
  = \sum_{n,n' \in \TwoP}
         (n'-n) N^n \NBar^\nBar \, \widetilde{\SS}(n,n') \,.
$$
In general, the matrix $\widetilde{\SS}$ may depend on node and/or
cycle, even when the gas is deterministic.  For example, it may take
on alternate values at odd and even cycles, or at adjacent nodes, or
both.  Since $\widetilde{\SS}$ represents the collision map (every
state $n$ must have exactly one image $n'$), it satisfies
\begin{equation}
  \label{map}
  \sum_{n'\in\TwoP} \widetilde{\SS}(n,n') = 1 \,, \qquad
  \hbox{for every $n\in\TwoP$}.
\end{equation}
It is also clear that the collision map is one-to-one if and only if
$\widetilde{\SS}$ satisfies
\begin{equation}
  \label{inv}
  \sum_{n\in\TwoP} \widetilde{\SS}(n,n') = 1 \,, \qquad
  \hbox{for every $n'\in\TwoP$}.
\end{equation}

  For simplicity, we will model all lattice gases as time stationary,
spatially homogeneous stochastic processes.  Let $\SS$ be the expected
value of $\widetilde{\SS}$.  Then
$$
  \widetilde{\SS}(n,n')
  = \left\{ \begin{array}{ll}
            1 \,, & \hbox{ with probability $\SS(n,n')$} \,; \\
            0 \,, & \hbox{ with probability $1-\SS(n,n')$} \,.
            \end{array} \right.
$$
Of course, if the gas has only one possible collision map
$\widetilde{\SS}$, then $\SS=\widetilde{\SS}$.  The
$2^{|\P|}\times 2^{|\P|}$ matrix $\SS$ is called the
{\bf local transition matrix} of the lattice gas.  By (\ref{map}),
$\SS$ satisfies
\begin{equation}
  \label{uni}
  \sum_{n'\in\TwoP} \SS(n,n') = 1 \,, \qquad
  \hbox{for every $n\in\TwoP$}.
\end{equation}
It is also clear from (\ref{inv}) that if every possible collision map
is one-to-one then $\SS$ also satisfies
\begin{equation}
  \label{rev}
  \sum_{n\in\TwoP} \SS(n,n') = 1 \,, \qquad
  \hbox{for every $n'\in\TwoP$}.
\end{equation}

  Let $\widetilde{F}=\hbox{\rm ExpVal}(N)$ where $N$ is determined
from the microdynamical equation (\ref{lgm}).  Then $\widetilde{F}$
takes on its value in the $|\P|$-dimensional interval $\II =[0,1]^\P$.
We consider the equation
\begin{equation}
  \label{lbm}
  \AA F = F + \CC(F) \,,
\end{equation}
where
\begin{equation}
  \label{collision}
  \CC(F) = \sum_{n,n'\in\TwoP}
                (n'-n) F^n \FBar^\nBar \, \SS(n,n') \,.
\end{equation}
Here $\CC$ is called a {\bf Boltzmann collision operator}.  If the
expected value operation passes through the nonlinearities of the
collision operator $\widetilde{\CC}$, of the lattice gas, then
$F=\widetilde{F}$.  Clearly, $\AA(F) = F+\CC(F) \in \II$.  Equation
(\ref{lbm}) can be viewed as a finite difference equation whose
solutions are grid functions $F(m,\i,\cdot)$ where
$F(0,\i,\cdot)=F_0(\i,\cdot)$ is the initial condition.  This
equation is called the {\bf lattice Boltzmann equation} for the
lattice gas automaton
\cite{BL89a,BL89b,CDDEFT90,CHL91,Elton90,ELR90,FHP86,FHHLPR87,HPP76,HP72,HPP73,Levermore92,LCCDLR91}.


\section{Boltzmann Collision Operators}

  We examine the relationship between the concepts of conservation,
equilibria and dissipation for a Boltzmann collision operator $\CC$
given by (\ref{collision}).  These relations are not special to
convective-diffusive lattice gases \cite{Doolen91}, but rather very
general.  The discussion here emphasizes this generality and closely
parallels the treatment in \cite{BGL91} of Boltzmann collision
operators without exclusion terms.

  The sum of any scalar or vector valued function $f=f(p)$ over
the variable $p$ will be denoted by $\< f\>$:
$$
  \< f \> = \sum_{p\in\P} f(p) \,.
$$

  Concepts of conservation are central to the existence of macroscopic
limits.  Two of them appear below which will be shown to be equivalent
for collision operators satisfying suitable conditions.

\begin{definition}
A mapping $e\,:\,\P\to\R$ (alternatively a vector $e\in\R^\P$) is said
to be a {\bf locally conserved quantity} for the collision operator
$~\CC$ if $\<e\,\CC(F)\> = 0$, for every $F \in \II$.
\end{definition}

\noindent
Note that a locally conserved quantity $e$ for the operator
$\CC$ given by (\ref{collision}) satisfies
$$
  \sum_{n,n' \in \TwoP}
  \< (n'-n) e \> F^n \FBar^\nBar \, \SS(n,n') = 0 \,,
$$
for every $F\in\II$.  Since the family of polynomials
parameterized by $n$ and given by $F \mapsto F^n \FBar^\nBar$ is
linearly independent, the above equality holds if and only if the
coefficient of each $F^n \FBar^\nBar$ vanishes:
\begin{equation}
  \label{eq7}
  \sum_{n' \in \TwoP} \< (n'-n) e \> \, \SS(n,n') = 0 \,,
\end{equation}
for every $n \in \TwoP$.

  Another notion of conservation is one that holds for individual
collisions.

\begin{definition}
A vector $e\in\R^\P$ is said to be a {\bf microscopically conserved
quantity} for the collision operator $\CC$ given by (\ref{collision})
if
\begin{equation}
  \label{eq8}
  \< (n'-n) e \> \SS(n,n') = 0 \,,
\end{equation}
for every $n,n'\in\TwoP$.
\end{definition}

\noindent
While it is clear from comparing (\ref{eq7}) and (\ref{eq8}) that any
microscopically conserved quantity is also locally conserved, the
converse is generally not true.

  The converse is true for the following class of collision operators,
however.

\begin{definition}
The operator $\CC$ is said to be in {\bf detailed balance} if
\begin{equation}
  \label{bal}
  \SS(n',n) = \SS(n,n') \,, \qquad
  \hbox{for every $n,n'\in\TwoP$} \,,
\end{equation}
and in {\bf semidetailed balance} if (see (\ref{uni}) and (\ref{rev}))
\begin{equation}
  \label{semi}
  \sum_{n' \in \TwoP} \SS(n',n)
  = \sum_{n' \in \TwoP} \SS(n,n') = 1 \,, \qquad
  \hbox{for every $n\in\TwoP$} \,.
\end{equation}
\end{definition}

\noindent
Clearly, the notion of semidetailed balance is a weakening of the
detailed balance condition.  Its usefulness arises through the
following characterization, the proof of which is immediate.

\begin{lemma} \label{lem3.1}
The collision operator $\CC$ given by (\ref{collision}) is in
semidetailed balance if and only if
\begin{equation}
  \label{eq6}
  \sum_{n,n' \in \TwoP} \nu(n) \, \SS(n,n')
  = \sum_{n,n' \in \TwoP} \nu(n') \, \SS(n,n') \,,
\end{equation}
for every $\nu\,:\,\TwoP\to\R$.
\end{lemma}

  The first implication of semidetailed balance is the following.

\begin{theorem} \label{thm3.1}
For any collision operator $\CC$ given by (\ref{collision}) that
is in semidetailed balance, any locally conserved quantity is also
microscopically conserved.
\end{theorem}

\begin{proof}
Let $e\in\R^\P$ be a locally conserved quantity.  Multiplying
(\ref{eq7}) by $\<ne\>$ and summing over $n$ gives
\begin{eqnarray}
  \label{eq9}
  0 & = & \sum_{n,n'\in\TwoP}
               \< (n'-n) e \> \< n e \> \, \SS(n,n')
\\
  \nonumber
  & = & - \sum_{n,n'\in\TwoP}
               \big( \< n e \>^2 - \< n' e \> \< n e \> \big)
               \, \SS(n,n') \,.
\end{eqnarray}
Applying (\ref{eq6}) of Lemma \ref{lem3.1} to half of the first term
inside of the last sum of (\ref{eq9}) gives
\begin{eqnarray*}
  0 & = & \sum_{n,n'\in\TwoP}
               \big( \tfrac12 \< n e \>^2
                   + \tfrac12 \< n' e \>^2
                   - \< n' e \> \< n e \> \big) \, \SS(n,n')
\\
    & = & \sum_{n,n'\in\TwoP}
               \tfrac12 \< (n-n') e \>^2 \SS(n,n') \,.
\end{eqnarray*}
Since each term of this last sum is nonnegative then all of them must
be equal to zero.  But that implies (\ref{eq8}) is satisfied and
shows that $e$ is also microscopically conserved.
\end{proof}

  The set of all locally conserved quantities of $\CC$ is a linear
subspace of $\R^{\P}$ denoted by $\E$ and assumed to be nontrivial.
Let $K$ be the dimension of $\E$ and $\{e_j\,|\,j=1,\ldots,K\}$
be a basis.  Let $\evec:\P\to\R^K$ where the components of $\evec(p)$
are these basis vectors.  Vectors in $\R^K$ will be denoted with arrows.
The Euclidean inner product of two such vectors, $\alphavec$ and
$\betavec$, is denoted by $\alphavec \odot \betavec$.

  Boltzmann collision operators in semidetailed balance also satisfy
the following $H$-theorem.

\begin{theorem}[$H$-Theorem] \label{H-thm}
Suppose the collision operator $\CC$ given by (\ref{collision}) is in
semidetailed balance.  Then it has the {\bf dissipation property}
\begin{equation}
  \label{dis}
  \< \CC(F) \log(F/\FBar) \> \leq 0 \,,
\end{equation}
for every $F\in\II$.  Moreover, the
following characterizations of equilibrium are equivalent:
\begin{eqnarray}
  \nonumber
  \hbox{  (i)}~ & \CC(F) = 0 \,,
\\
  \label{equil}
  \hbox{ (ii)}~ & \<\CC(F)\log(F/\FBar)\> = 0 \,,
\\
  \nonumber
  \hbox{(iii)}~ & F^n \FBar^\nBar = F^{n'} \FBar^{\nBar'} \,, \quad
                  \forall\, n,n' \in \TwoP
                  \hbox{ such that $\SS(n,n')>0$} \,,
\\
  \nonumber
  \hbox{ (iv)}~ & F = M(\betavec)
                  \hbox{ for some $\betavec\in\R^K$} \,,
\end{eqnarray}
where $M(\betavec)$ is given by,
\begin{equation}
  \label{max}
  M(\betavec)
  = \frac{1}{1+\exp(-\betavec \odot \evec)} \,.
\end{equation}
Here $\evec$ is a basis of $\E$, the $K$-dimensional space of
locally conserved quantities.
\end{theorem}

\noindent
{\bf Remark:}  The form of the local equilibrium given in (\ref{max})
is that of a Fermi-Dirac density, which is the quantum mechanical
analogue of the classical Maxwellian density for particles satisfying
an exclusion principle, hence the designation $M$.

\begin{proof}
Since the logarithm of a product is the sum of its logarithms, one can
verify that
\begin{equation}
  \label{eq11}
  \< (n-n') \log(F/\FBar) \>
  = - \log\!\left( \frac{F^{n'} \FBar^{\nBar'}}
                        {F^n \FBar^\nBar} \right) \,.
\end{equation}
Since $\CC$ is in semidetailed balance, letting
$\nu(n)=F^n\FBar^\nBar$ in (\ref{eq6}) of Lemma \ref{lem3.1} yields
$$
  0 = \sum_{n,n' \in \TwoP} F^{n'} \FBar^{\nBar'} \SS(n,n')
      - \sum_{n,n' \in \TwoP} F^n \FBar^\nBar \SS(n,n') \,.
$$
Hence,
\begin{eqnarray}
  \label{eqn12}
  - \<\CC(F) \log(F/\FBar)\> \!\!
  & = & \!\! \sum_{n,n' \in \TwoP} \<(n-n') \log(F/\FBar)\>
                  F^n \FBar^\nBar \SS(n,n') \\
  \nonumber
  & = & \!\! \sum_{n,n' \in \TwoP} \Bigg(
                 \frac{F^{n'}\FBar^{\nBar'}}{F^n \FBar^\nBar}-1
                 - \log\!\Bigg( \frac{F^{n'} \FBar^{\nBar'}}
                                     {F^n \FBar^\nBar} \Bigg) \!
             \Bigg) F^n \FBar^\nBar \SS(n,n') \,.
\end{eqnarray}
Since $y-1-\log y\geq 0$ for every $y\in\R_+$, every term in the last
sum of (\ref{eqn12}) is nonnegative, so that the collision operator
$\CC$ satisfies the dissipation property.

  The characterization of equilibria (\ref{equil}) is argued as
follows: (i) implies (ii) implies (iii) implies (iv) implies (i).
The first implication is obvious.  Assuming (ii), the last sum in
(\ref{eqn12}) is zero and each of its nonnegative terms must vanish.
This gives the formula
$$
  \left( F^{n'}\FBar^{\nBar'} - {F^n \FBar^\nBar}
  - {F^n \FBar^\nBar}
    \log\!\left( \frac{F^{n'} \FBar^{\nBar'}}
                      {F^n \FBar^\nBar} \right) \right)
  F^n \FBar^\nBar \SS(n,n') = 0 \,,
$$
for every $n,n'\in\TwoP$.  Since $(y,z)\mapsto y-z-z\log(y/z)$ over
$\R_+\times\R_+$ is nonnegative, and vanishes only on the diagonal
$y=z$, it then follows that
$$
  \big( F^n \FBar^\nBar - F^{n'} \FBar^{\nBar'} \big)
                                            \SS(n,n') = 0 \,,
$$
for every $n,n'\in\TwoP$, which gives (iii).  Assuming (iii) then
(\ref{eq11}) implies
$$
  \<(n-n') \log(F/\FBar)\> \SS(n,n')
  = - \log\!\left( \frac{F^{n'} \FBar^{\nBar'}}
                        {F^n \FBar^\nBar} \right) \SS(n,n') = 0 \,,
$$
for every $n,n' \in \TwoP$.  Thus $\log(F/\FBar)$ satisfies
(\ref{eq8}) and is therefore a microscopically conserved quantity.
Hence,
$$
  \log(F/\FBar) = \sum_{j=1}^K \beta_j e_j
                  = \betavec \odot \evec
$$
for some vector $\betavec=(\beta_1,\ldots,\beta_K)^T\in\R^K$.
Solving this for $F$ yields (iv).  Finally, assuming (iv) and using
the fact that all locally conserved quantities are microscopically
conserved (Theorem \ref{thm3.1}) and employing (\ref{eq6}) of Lemma
\ref{lem3.1}, it is easy to show that $\CC(M(\betavec))=0$ for every
$\betavec\in\R^K$.
\end{proof}

  Finally, another consequence of the property of semidetailed balance
is the characterization of the Fredholm alternative for the first
derivative of $\CC$ evaluated at any given local equilibrium
$M=M(\betavec)$.  The {\bf linearized collision operator} at $M$ is
$\LL_\M$ defined by
\begin{equation}
  \label{deriv}
  \LL_\M h \equiv \DD\CC(M) h
  \equiv \del_\eps \CC(M + \eps h) \Big|_{\eps=0} \,,
\end{equation}
for every $h \in \II$.  First observe that every $e\in\E$ can be
written as $e=\alphavec\odot\evec$ for some unique $\alphavec\in\R^K$.
The formula for the local equilibria (\ref{max}) then yields
\begin{equation}
  \alphavec \odot \del_\betavec M(\betavec)
  = M(\betavec) (1 - M(\betavec)) \alphavec \odot \evec
  = M(\betavec) (1 - M(\betavec)) e \,.
\end{equation}
Defining $W_\M=W_\M(\betavec)\equiv M(\betavec)(1-M(\betavec))$, one
sees that
\begin{equation}
  0 = \alphavec \odot \del_\betavec \, \CC(M(\betavec))
    = \DD\CC(M(\betavec))
      \big( \alphavec \odot \del_\betavec M(\betavec) \big)
    = \DD\CC(M(\betavec)) W_\M(\betavec) e \,,
\end{equation}
which implies that $\E\subset\NULL(\LL_\M W_\M)$.

  Next, let $\LL_\M^T$ denote the transpose of $\LL_\M$
defined by
\begin{equation}
  \label{trans}
  \< h \LL_\M^T g \> = \< g \LL_\M h \> \,, \qquad
  \hbox{for every $h,g \in \II$} \,.
\end{equation}
If $e$ is a locally conserved quantity of $\CC$ then
$$
  \< e \, \LL_\M h \>
  = \del_\eps \< e \CC(M + \eps h) \> \Big|_{\eps=0} = 0 \,,
$$
for every $h \in \II$, which by (\ref{trans}) implies that
$\E\subset\NULL(\LL_\M^T)$.

  The above inclusions become equalities for collision operators that
are in semidetailed balance.

\begin{theorem} \label{thm3.3}
If the collision operator $\CC$ given by (\ref{collision}) is in
semidetailed balance and $M$ is any local equilibrium of $\CC$, then
its linearization $\LL_\M$ defined by (\ref{deriv}) satisfies
$$
  \E = \NULL(\LL_\M W_\M) = \NULL(\LL_\M^T)
     = \NULL(\LL_\M W_\M + W_\M \LL_\M^T) \,,
$$
and moreover, $\< g \, \LL_\M W_\M g \> < 0$ for every $g\notin\E$.
\end{theorem}

\begin{proof}
Above it was shown that $\E$ is contained in both $\NULL(\LL_\M W_\M)$
and $\NULL(\LL_\M^T)$.  Since every $g\in\R^\P$ satisfies
$$
  \< g \, \LL_\M W_\M g \>
  = \tfrac12 \< g \, (\LL_\M W_\M + W_\M \LL_\M^T) g \>
  = \< g \, W_\M \LL_\M^T g \> \,,
$$
it is clear that $\NULL(\LL_\M^T)$ and $\NULL(\LL_\M W_\M)$ are each
contained in $\NULL(\LL_\M W_\M + W_\M \LL_\M^T)$.  All that remains
to be shown is that $\NULL(\LL_\M W_\M + W_\M \LL_\M^T)\subset\E$.  A
direct calculation following (\ref{deriv}) and using $\CC(M)=0$ yields
\begin{equation}
  \label{lin}
  \LL_\M W_\M g = \sum_{n,n' \in \TwoP}
                  (n' - n) \< n \, g \> M^n \MBar^\nBar \SS(n,n') \,.
\end{equation}
If $g\in\NULL(\LL_\M W_\M + W_\M \LL_\M^T)$ then using semidetailed
balance (as in the proof of Theorem \ref{thm3.1}) shows that
\begin{eqnarray*}
  0 = - \< g \, \LL_\M W_\M g \>
  & = & - \sum_{n,n' \in \TwoP}
              \< (n' - n) g \> \< n \, g \> M^n \MBar^\nBar \SS(n,n')
\\
  & = & \sum_{n,n' \in \TwoP}
            \tfrac12 \< (n' - n) g \>^2 M^n \MBar^\nBar \SS(n,n') \,.
\end{eqnarray*}
Since each term of this last sum is nonnegative, all of them must
be equal to zero.  But that means $g$ satisfies (\ref{eq8}) and is
therefore a locally conserved quantity ($g\in\E$).  Moreover, it is
clear that the sum is zero if and only if $g\in\E$.
\end{proof}

\noindent
An immediate consequence of Theorem \ref{thm3.3} is the Fredholm
alterative that for any $f\in\R^\P$ the overdetermined system
\begin{equation}
  \label{pseudo}
  \LL_\M h = f \,, \qquad \< \evec \, h \> = 0 \,,
\end{equation}
has a solution if and only if $\< \evec f \>=0$, in which case the
solution is unique and is denoted by $\LL_M^{-1}f$.  The operator
$\LL_\M^{-1}$ is a (left) {\bf pseudoinverse} of $\LL_\M$.


\section{The Hilbert Expansion and Diffusion}

  Here we give the characterization of convective-diffusive lattice
gases by properties of their Boltzmann approximations, more precisely,
by properties of their local equilibria.  The notion of the continuum
limit of such a gas involves refining the lattice domain $\X$ within
the macroscopic domain $\Omega\in\RD$ and is formulated in terms of
the vanishing of a parameter $\delta>0$ that is related to lattice
spacing $\Delta x$ and time cycle interval $\Delta t$ by
\begin{equation}
  \label{scal}
  \Delta x = \delta \, L \,, \qquad
  \Delta t = \delta^2 T \,,
\end{equation}
where $L,T>0$ are macroscopic length and time scales.  Of course, the
scaling of $\Delta t=O((\Delta x)^2)$ is the usual diffusive scaling,
but not every lattice gas has macroscopic dynamics that is consistent
with it.

  Here we consider Boltzmann collision operators of the form
\begin{equation}
  \label{czero}
  \CC = \CC^{(0)} + \delta \, \CC^{(1)} \,,
\end{equation}
where $\CC^{(0)}$ and $\CC^{(1)}$ are Boltzmann operators such that
every locally conserved quantity of $\CC^{(0)}$ is also locally
conserved by $\CC^{(1)}$.  The spaces of locally conserved quantities
of $\CC^{(0)}$ and $\CC$ therefore coincide and we denote this space
by $\E$ and let $\evec$ denote a basis.  Moreover, we assume that
$\CC^{(0)}$ is in semidetailed balance; hence its equilibria are
given by $F=M(\betavec)$, and it satisfies the $H$-Theorem.  Finally,
we assume that the lattice gas is diffusive:

\begin{definition} \label{diffusive}
A lattice gas such as given above is called {\bf diffusive} provided
\begin{equation}
  \label{diff}
  \< \v \evec M(\betavec) \> = 0 \,, \qquad
  \hbox{for every $\betavec\in\R^K$} \,.
\end{equation}
\end{definition}

\noindent
This condition will insure that the time scale of the macroscopic
dynamics will be consistent with the diffusion scaling (\ref{scal}).

  The limiting convective-diffusive macroscopic dynamics of the gas is
established as follows.  First, a family of approximate solutions of
the lattice Boltzmann equation parametrized by $\delta$ is constructed
from smooth functions over the $(t,\x)$-domain $\R_+\times\Omega$ that
are solutions of convection-diffusion equations.  Then it is shown
that the exact solution of the lattice Boltzmann equation and the
approximate solution converge in some sense.  The first step, carried
out below, is the lattice Boltzmann version of the consistency step of
most numerical convergence proofs, while the second will follow from a
stability argument given in the next section.

  Given any function $H=H(t,\x,p)$ that is a smooth mapping from the
$(t,\x)$-domain $\R_+\times\Omega$ into $\R_+^\P$, the Taylor expansion
of $\AA H(t,\x,p)$ about $(t,\x)$ is
\begin{eqnarray}
  \nonumber
  \AA H(t,\x,p)
  & = & H(t+\delta^2 T,\x + \delta L \v(p),p)
\\
  \label{advect}
  & = & \sum_{j=0}^\infty
        \frac{1}{j!}
        \big[ \delta^2 T \del_t
            + \delta \,L \v(p) \cdot \GRAD \big]^j H(t,\x,p) \,.
\end{eqnarray}
Grouping terms by order of $\delta$ gives
\begin{eqnarray}
  \nonumber
  \AA H - H =
  &   & \delta \,L (\v \cdot \GRAD) H
\\
  \nonumber
  & + & \delta^2 \big[ T \del_t
                     + \tfrac12 L^2 (\v \cdot \GRAD)^2 \big] H
\\
  \label{As}
  & + & \delta^3 \big[ T L \del_t (\v \cdot \GRAD)
                     + \tfrac16 L^3 (\v \cdot \GRAD)^3 \big] H
\\
  \nonumber
  & + & \delta^4 \big[ \tfrac12 T^2 \del_t^2
                     + \tfrac12 T L^2 \del_t (\v \cdot \GRAD)^2
                     + \tfrac1{24} L^4 (\v \cdot \GRAD)^4 \big] H
\\
  \nonumber
  & + & \cdots \,.
\end{eqnarray}

  We construct an approximate solution to the lattice Boltzmann
equation
\begin{equation}
  \label{approx}
  \AA H - H = \CC(H) \,
\end{equation}
by formally expanding $H$ in powers of $\delta$ as
\begin{equation}
  \label{series}
  H(t,\x,p) = \sum_{k=0}^\infty \delta^k h^{(k)}(t,\x,p) \,.
\end{equation}
This series is the lattice analogue of the classical
{\bf Hilbert expansion} of kinetic theory through which one formally
passes to the limiting macroscopic dynamics.  Notice that, as with
classical Hilbert expansions, the advection side (\ref{As}) is
$O(\delta)$.

  Expanding the left side of (\ref{approx}) in powers of $\delta$
gives
\begin{eqnarray}
  \label{Taylor1}
  \AA H - H
  & = & (\AA - I) \sum_{k=0}^\infty \delta^k h^{(k)}
\\
  \nonumber
  & \equiv & a^{(0)} + \delta \, a^{(1)}
                     + \delta^2 a^{(2)}
                     + \delta^3 a^{(3)} + \cdots \,,
\end{eqnarray}
where
\begin{eqnarray*}
  a^{(0)} & = & 0 \,,
\\
  a^{(1)} & = & L (\v \cdot \GRAD) h^{(0)} \,,
\\
  a^{(2)} & = & \big[ T \del_t
                     + \tfrac12 L^2 (\v \cdot \GRAD)^2 \big] h^{(0)}
                + L (\v \cdot \GRAD) h^{(1)} \,,
\\
  a^{(3)} & = & \big[ T L \del_t (\v \cdot \GRAD)
                     + \tfrac16 L^3 (\v \cdot \GRAD)^3 \big] h^{(0)}
\\
          &   & + \, \big[ T \del_t
                      + \tfrac12 L^2 (\v \cdot \GRAD)^2 \big] h^{(1)}
                + L (\v \cdot \GRAD) h^{(2)} \,,
\\
          & \vdots & \,.
\end{eqnarray*}
Notice that $a^{(j)}$ depends on $h^{(0)}$ through $h^{(j-1)}$, but
not on $h^{(j)}$.

  Similarly expanding the right side of (\ref{approx}) gives
\begin{eqnarray}
  \label{Taylor2}
  \CC(F)
  & = & \CC^{(0)}\bigg( \sum_{k=0}^\infty \delta^k h^{(k)} \bigg)
        + \delta \,
          \CC^{(1)}\bigg( \sum_{k=0}^\infty \delta^k h^{(k)} \bigg)
\\
  \nonumber
  & \equiv & c^{(0)} + \delta \, c^{(1)}
                     + \delta^2 c^{(2)}
                     + \delta^3 c^{(3)} + \cdots \,,
\end{eqnarray}
where
\begin{eqnarray*}
  c^{(0)} & = & \CC^{(0)}(h^{(0)}) \,,
\\
  c^{(1)} & = & \DD\CC^{(0)}(h^{(0)}) \cdot h^{(1)}
                + \CC^{(1)}(h^{(0)}) \,,
\\
  c^{(2)} & = & \DD\CC^{(0)}(h^{(0)}) \cdot h^{(2)}
                + \tfrac12
                  \DD^2\CC^{(0)}(h^{(0)}) \cdot h^{(1)}h^{(1)}
                + \DD\CC^{(1)}(h^{(0)}) \cdot h^{(1)} \,,
\\
  c^{(3)} & = & \DD\CC^{(0)}(h^{(0)}) \cdot h^{(3)}
                + \DD^2\CC^{(0)}(h^{(0)}) \cdot h^{(1)}h^{(2)}
\\
          &   & + \, \tfrac16
                  \DD^3\CC^{(0)}(h^{(0)}) \cdot h^{(1)}h^{(1)}h^{(1)}
\\
          &   & + \, \DD\CC^{(1)}(h^{(0)}) \cdot h^{(2)}
                + \tfrac12
                  \DD^2\CC^{(1)}(h^{(0)}) \cdot h^{(1)}h^{(1)} \,,
\\
          & \vdots & \,.
\end{eqnarray*}
The general form for the $c^{(j)}$ in (\ref{Taylor2}) is
\begin{equation}
  \label{rj}
  c^{(j)} = \DD\CC^{(0)}(h^{(0)}) \cdot h^{(j)} + r^{(j)} \,,
\end{equation}
where $r^{(j)}$ refers to all the remaining terms of $c^{(j)}$, each
of which depends on $h^{(0)}$ through $h^{(j-1)}$, but not on $h^{(j)}$.
Notice that since $r^{(j)}$ is just the sum of derivatives of the
collision operator $\CC$, it automatically satisfies
$\<\evec\,r^{(j)}\>=0$.

  Matching (\ref{Taylor1}) to (\ref{Taylor2}) at leading order gives
\begin{equation}
  \label{order0}
  \CC^{(0)}(h^{(0)}) = 0 \,,
\end{equation}
which by the $H$-Theorem implies that
\begin{equation}
  \label{h0}
  h^{(0)} = M = M(\betavec) \,, \qquad
  \hbox{for some $\betavec \in \R^K$} \,.
\end{equation}
Here $\betavec=\betavec(t,\x)$ is a smooth function still to be
determined.

  Matching (\ref{Taylor1}) to (\ref{Taylor2}) order by order for $j>0$
gives a linear equation for $h^{(j)}$ in the form
\begin{equation}
  \label{orderj}
  \LL_\M h^{(j)} = a^{(j)} - r^{(j)} \,,
\end{equation}
where the right side depends on $h^{(0)}$ through $h^{(j-1)}$, but not
on $h^{(j)}$.  Being of the form (\ref{pseudo}), the linear equation
(\ref{orderj}) will have a solution if and only if its right side
satisfies the solvability condition
$$
  0 = \< \evec (a^{(j)} - r^{(j)}) \> \,.
$$
Since $r^{(j)}$ automatically satisfies $\<\evec\,r^{(j)}\>=0$, this
condition reduces to
\begin{equation}
  \label{solvej}
  0 = \< \evec \, a^{(j)} \> \,.
\end{equation}
This satisfied, the general solution is then
\begin{equation}
  \label{hj}
  h^{(j)} = \LL_\M^{-1} (a^{(j)} - r^{(j)})
            + W_\M \betavec^{(j)} \odot \evec \,, \qquad
  \hbox{for some $\betavec^{(j)} \in \R^K$} \,,
\end{equation}
where $\LL_\M^{-1}$ is the left pseudoinverse of $\LL_\M$.  Here
$\betavec^{(j)}=\betavec^{(j)}(t,\x)$ is a smooth function to be
determined.  It is this arbitrariness in the solution of $h^{(j)}$
at order $j$ that allows exactly the freedom necessary to impose the
solvability condition (\ref{solvej}) at order $j+2$ of the matching
procedure.

  In particular, the leading order $\betavec$ of (\ref{h0}) will be
determined by the solvability condition (\ref{solvej}) at order $2$.
Indeed, at order $1$, (\ref{orderj}) becomes
\begin{equation}
  \label{order1}
  \LL_\M h^{(1)}
  = L (\v \cdot \GRAD) M - \CC^{(1)}(M) \,.
\end{equation}
Its solvability condition (\ref{solvej}) is simply
\begin{equation}
  \label{solve1}
  0 = \< \evec \v \cdot \GRAD M \>
    = \DIV \< \v \evec M \> \,,
\end{equation}
which is automatically satisfied since the right side vanishes
identically when the lattice gas is diffusive (\ref{diff}).  Solving
(\ref{order1}) for $h^{(1)}$ then yields
\begin{equation}
  \label{h1}
  h^{(1)} = \LL_\M^{-1} \big( L \v \cdot \GRAD M - \CC^{(1)}(M) \big)
            + W_\M \betavec^{(1)} \odot \evec \,,
\end{equation}
for some $\betavec^{(1)}\in\R^K$.  At order $2$,
(\ref{orderj}) becomes
\begin{eqnarray}
  \label{order2}
  \LL_\M h^{(2)}
  & = & \big[ T \del_t
              + \tfrac12 L^2 (\v \cdot \GRAD)^2 \big] M
         + L (\v \cdot \GRAD) h^{(1)}
\\
  \nonumber
  &   &  - \tfrac12 \DD^2\CC^{(0)}(M) \cdot h^{(1)} h^{(1)}
         - \DD\CC^{(1)}(M) \cdot h^{(1)} \,.
\end{eqnarray}
Its solvability condition (\ref{solvej}) is
\begin{eqnarray}
  \label{solve2}
  0 & = & \< \evec \, \big[ T \del_t M
                         + \tfrac12 L^2 (\v \cdot \GRAD)^2 M
                         + L (\v \cdot \GRAD) h^{(1)} \big] \>
\\
  \nonumber
    & = & T \del_t \< \evec M \>
          + L^2 \DIV \< (\v \evec (\LL_\M^{-1} + \tfrac12 I)
                         \v \cdot \GRAD M \>
          - L \DIV \< \evec \v \CC^{(1)}(M) \> \,,
\end{eqnarray}
which leads to the evolution equation for $\betavec=\betavec(t,\x)$
\begin{equation}
  \label{evol}
  \del_t \< \evec M \>
  = \frac{L}{T} \DIV \< \evec \v \CC^{(1)}(M) \>
    - \frac{L^2}{T}
      \DIV \big( \< (\v \evec (\LL_\M^{-1} + \tfrac12 I)
                    W_\M \evec \v \cdot \GRAD \>
                 \odot \betavec \big) \,,
\end{equation}
where $M=M(\betavec)$.  The first term on the right is a convection
term provided it is nonzero.  This will be the case whenever $M$ is
{\it not\/} a local equilibrium of $\CC^{(1)}$, which can only happen
if $\CC^{(1)}$ is {\it not\/} in semidetailed balance (recall the
$H$-Theorem).  The second term on the right is a diffusion term
provided the diffusion 4-tensor
\begin{equation}
  \label{tensor}
  \< (\v \evec (\LL_\M^{-1} + \tfrac12 I) W_\M \evec \v \>
\end{equation}
is negative definite.  The negativity of
$\< (\v \evec \LL_\M^{-1} W_\M \evec \v \>$ follows directly from
Theorem \ref{thm3.3} and the fact that $W_\M\evec\v$ is not in $\E$ since
(\ref{diff}) implies that $\<W_\M\evec\,\evec\v\>=0$.  That this
negativity is enough to overcome the antidiffusion term in
(\ref{tensor}) that arises from the second term in the Taylor
expansion of the discrete advection is a deeper fact due to Henon
\cite{FHHLPR87}.  This will be made explicit in the specific numerical
examples that we study later.

  In general, the determination of $\betavec^{(j)}$ at order $j+2$ is
a consequence of the diffusive property (\ref{diff}) of $\CC$ since,
for $j>0$, it is seen that $a^{(j+1)}$ has the general form
\begin{equation}
  a^{(j+1)} = L (\v \cdot \GRAD) W_\M \betavec^{(j)} \odot \evec
              + s^{(j+1)} \,,
\end{equation}
where $s^{(j+1)}$ denotes the remaining terms, each of which depends
on $h^{(0)}$ through $h^{(j-1)}$, but not on $\betavec^{(j)}$.
Differentiating the diffusive property (\ref{diff}) with respect to
$\betavec$ leads to the identity
\begin{equation}
  \label{diffderiv}
  \< \v \evec\, W_\M(\betavec) \betavec^{(j)} \odot \evec \>
  = \betavec^{(j)} \odot \del_\betavec \< \v \evec M(\betavec) \>
  = 0 \,.
\end{equation}
The solvability condition (\ref{solvej}) at order $j+1$ then reduces to
\begin{eqnarray}
  \nonumber
  0 & = & \< \evec \, a^{(j+1)} \>
\\
    & = & L \DIV \< \v \evec\, W_\M \betavec^{(j)} \odot \evec \>
          + \< \evec \, s^{(j+1)} \>
\\
  \nonumber
    & = & \< \evec \, s^{(j+1)} \> \,.
\end{eqnarray}
This gives a forced, linearization of the convection-diffusion
equation (\ref{evol}) that governs the evolution of the as yet
undetermined $\betavec^{(j-1)}$.  In this way one can systematically
construct $h^{(k)}$ order by order from solutions of (\ref{evol}) and
its forced linearizations.


\section{Consistency, Stability and Convergence}

  We consider finite truncations of the formal expansion constructed
in the last section
\begin{equation}
  \label{series2}
  H^{(q)}(t,\x,\p) = \sum_{k=0}^q \delta^k h^{(k)}(t,\x,\p) \,.
\end{equation}
By construction, $H^{(q)}(m\delta^2 T,\i\delta L,p)$ satisfies
$$
  \AA H^{(q)} - H^{(q)} - \CC[H^{(q)}] = \TT^{(q)} \,,
$$
where formally $\TT^{(q)}=O(\delta^{q+1})$.

\begin{definition}
Let $q>0$ be a fixed integer and $B_\delta$ a finite dimensional
Banach space with ${\ell}_1$-norm $\| \cdot \|_\delta$.

\noindent
(i) {\bf Consistency }

If
$$
  \lim_{\Delta t \to 0}
       \frac{1}{\Delta t}
       \|\TT^{(q)}(m, \cdot ,\cdot)\|_{\delta } = 0 \,,
$$
then the lattice Boltzmann method is said to be {\bf consistent}.

\noindent
(ii) {\bf Convergence}

  If $F(m,\cdot,\cdot) \in B_{\delta }$ is the solution to the
lattice Boltzmann method (\ref{lbm}) and
$$
  \lim_{\Delta t \to 0}
       \|F(m,\cdot,\cdot) - H^{(q)}(m,\cdot,\cdot)\|_{\delta } = 0 \,,
$$
for all integers $m$ such that $0 \leq m \Delta t \leq T$, then the
lattice Boltzmann method is said to be {\bf convergent}.

\noindent
(iii) {\bf Stability}

  Define the block diagonal matrix
$L_{\delta }: B_{\delta } \to B_{\delta }$ where the $\i$-th
diagonal block is defined by
$$
  \LL(F,H^{(q)}) h
  = \int_0^1 \left[
              I + \DD\CC\big( (1-s) F + s H^{(q)} \big)
              \right] \cdot h \, ds \,,
$$
for $h\in\II$.  The lattice Boltzmann method is said to be {\bf stable}
if for some $\tau>0$, the family of matrices
$$
  \left\{ \prod_{k=0}^m L_{\delta } \,:\,
          \hbox{$0<\Delta t<\tau$ and $0\leq m\Delta t\leq T$}
  \right\} \,,
$$
is uniformly bounded.
\end{definition}

  We now prove a theorem that resembles the easy direction of the
classical Lax Equivalence Theorem found in most finite difference
texts, see \cite{RM67} for example.

\begin{theorem}
\label{thm51}
Suppose a lattice Boltzmann method is consistent.  Then stability is a
sufficient condition for convergence.
\end{theorem}

\begin{proof}
Let $F(m,\i,\cdot)$ be a solution to the lattice Boltzmann equation
(\ref{lbm}) and
$$
  E = F - H^{(q)} \,.
$$
Note that
$$
  \CC(F) - \CC(H^{(q)}) + E = \LL(F,H^{(q)}) E \,.
$$
Also,
$$
  \CC(F) - \CC(H^{(q)}) = \AA E - E + \TT^{(q)} \,,
$$
where
$$
  \lim_{\Delta t \to 0} \frac{1}{\Delta t}
       \|\TT^{(q)}(m,\cdot,\cdot)\|_{\delta } = 0 \,.
$$
Hence,
$$
  \LL(F,H^{(q)}) E = \AA E + \TT^{(q)} \,.
$$
There exists a permutation matrix $\PP$ such that
$$
  E(m+1,\cdot,\cdot) = \PP L_{\delta }\PP^t \PP\cdot E(m,\cdot,\cdot)
                       + \PP \TT^{(q)}(m,\cdot,\cdot) \,.
$$
Let $\widetilde{\TT}^{(q)} = \PP \TT^{(q)}(m,\cdot,\cdot)$ and
$\widetilde{L}_{\delta }= \PP L_{\delta }\PP^t$.  Then, by stability and
consistency, there exists a constants $C_1,C_2$ so that
\begin{eqnarray*}
  \| E(m+1,\cdot,\cdot)\|_{\delta }
  & = & \Big\| \sum_{n=0}^{m}
               \left( \prod_{r=n+1}^{m-1} \widetilde{L}_{\delta } \right)
               \widetilde{\TT}^{(q)}(n,\cdot,\cdot) \Big\|_{\delta }
\\
  & \leq &  C_1 \sum_{n=0}^{m-1}
                \| \TT^{(q)}(n,\cdot,\cdot) \|_{\delta }
              + \| \TT^{(q)}(m,\cdot,\cdot) \|_{\delta }
\\
  & \leq & (m-1) C_1 C_2 \Delta t + C_2 \Delta t
\\
  & \leq & m C \Delta t \,.
\end{eqnarray*}
Hence, the method is convergent.
\end{proof}

  We now establish sufficient conditions for stability of a lattice
Boltzmann method using the ideas of monotone difference methods, e.g.,
\cite{Sod88}.  Consider the operator $\BB$ defined on $\II$ having the
$p$-th coordinate function given by
\begin{equation}
  \BB[F](p) = F(p) + \CC[F](p) \,.
\end{equation}
The derivative of $\BB$ is the $|\P| \times |\P|$ Jacobian matrix
$\JJ_{\BB}(F)$ whose $p,q$-th entry is
$$
  \frac{\del}{\del F(q)} \BB[F](p) \,.
$$
We assume $\JJ_{\BB}(F)$ to be a continuous function of $F$.

\begin{definition}
  Let
$$
  \EE
  = \prod_{k=0}^{|\P|}~[M_-^{(k)},M_+^{(k)}]
  ~\subseteq~ \II \,,
$$
be a $|\P|$-dimensional interval upon which $\JJ_\BB$ is a nonnegative
matrix.  That is, $F\in\EE$ implies that each entry of $\JJ_{\BB}(F)$
is nonnegative.  Then $\EE$ is called a {\bf domain of monotonicity}
of the lattice Boltzmann method (\ref{lbm}).  The vectors
$$
  \MM_- = \left[ M_-(p) \right]_{p \in \P} \qquad \mbox{and} \qquad
  \MM_+ = \left[ M_+(p) \right]_{p \in \P}
$$
are called the {\bf extreme points} of $\EE$.
\end{definition}

  The following theorem demonstrates the invariance property of the
advection operator on a domain of monotonicity.

\begin{theorem} \label{thm52}
  Let $\EE$ be a domain of monotonicity for a lattice Boltzmann
method (\ref{lbm}) with extreme points $\MM_-$ and $\MM_+$.
If
$$
  \CC(\MM_-) = \CC(\MM_+) = 0 \,,
$$
then $\BB$ leaves $\EE$ invariant.
That is, $F \in \EE$ implies $\BB F \in \EE$.
\end{theorem}

\begin{proof}
Note that $\JJ_{\BB} \geq 0$ on $\EE$ implies the
coordinate function $\BB[F](p)$ is monotonically
increasing.  Moreover, continuity of $\JJ_{\BB}$
implies $\JJ_{\BB}(\MM_+) = 0$ so that $M_+(p)$
maximizes $\BB[F](p)$ on $[M_-(p),M_+(p)]$.  Hence,
$$
  \BB[F](p) \leq \BB[\MM_+](p)
                    = M_+(p) + \CC[M_+](p) =  M_+(p) \,.
$$
\end{proof}

A stability condition can be established for lattice Boltzmann methods
whose collision operators have
the following conservation property.

\begin{definition}
A collision operator $\CC$ is said to {\bf conserve mass} if
$\< \CC (F) \> = 0$ for all $F \in \II$. That is, $e \equiv 1$ is a
locally conserved quantity.
\end{definition}

\begin{theorem}
\label{thm53}
Let $\EE$ be any domain of monotonicity for the lattice
Boltzmann method (\ref{lbm}) such that the collision operator
is zero at the extreme points and suppose $F(0,\i,\cdot)$ and
$H^{(q)}(0,\i,\cdot)$ are vectors in $\EE$.  If $\CC$
conserves mass, then the method is stable.
\end{theorem}

\begin{proof}
Since $\JJ_{\BB}(F) \geq 0$ and the fact that
$\| \cdot\|_{\delta }$ is the $\ell_1$ norm, we get
\begin{eqnarray*}
  \| \JJ_{\BB}(F) \|_{\delta }
  & = & \max_{q\in\P} \bigg\{\<
             \frac{\del}{\del F(q)}[F +  \CC[F]] \> \bigg\}
\\
  & = & \max_{q\in\P}\bigg\{ 1 +  \<
             \frac{\del}{\del F(q)} \CC[F]\> \bigg\}
\\
  & = & \max_{q\in\P}\bigg\{ 1 + \frac{\del}{\del F(q)}
                                 \< \CC[F]\> \bigg\}
\\
  & = & 1 \,
\end{eqnarray*}
where the last equality follows from the fact that $\CC$ conserves
mass.  Since by Theorem \ref{thm52} $\BB$ leaves $\EE$ invariant, it
follows that $F(m,\i,\cdot)$ and $H^{(q)}(m,\i,\cdot)$ are both in
$\EE$ for all integers $m$.  Thus, the connectedness of $\EE$ implies
$$
  [F(m,\i,\cdot) - s(H(m,\i,\cdot) - F(m,\i,\cdot))]
  \in \EE \,, \qquad 0 \leq s \leq 1 \,.
$$
Hence for $G \in \II$ we get
\begin{eqnarray*}
  \| \LL(F,H^{(q)}) \cdot G \|_{\delta }
  & = & \Big\| \int_0^1
               \big[ I + \DD\CC\big( (1-s) F + s H^{(q)} \big) \big]
               \cdot G \, ds \Big\|_{\delta }
\\
  & \leq & \int_0^1
           \| I + \DD\CC\big( (1-s) F + s H^{(q)} \big) \|_{\delta }
           \cdot \|G\|_{\delta } \, ds
\\
  & \leq & \|G\|_{\delta } \int_0^1ds
\\
  & = & \| G\|_{\delta } \,.
\end{eqnarray*}
\end{proof}


\section{Example - A Nonlinear Diffusive System}

  We consider a lattice Boltzmann method, called LB1, constructed from
a lattice gas on a periodic square lattice. A particle can be in one
of $p=0,1,2,3$ possible states at a node, as are indicated in Table
\ref{LB1:states}.
\begin{table}[p]
\newcommand{\PICSINN}[5]{\setlength{\unitlength}{#5}
  \begin{picture}(0,0)\thicklines
    \put( 0.0,-0.5){#1(0,1){0.5}}
    \put(-0.5, 0.0){#2(1,0){0.5}}
    \put( 0.0, 0.5){#3(0,-1){0.5}}
    \put( 0.5, 0.0){#4(-1,0){0.5}}
  \end{picture}
}
\thinlines
\setlength{\unitlength}{4em}
\newlength{\Widths}
\setlength{\Widths}{0.8\unitlength}
\def\Entry#1#2#3#4{
        \begin{picture}(1,1)
                \put(0.5,0.5){#1}
        \end{picture}
        &
        \begin{picture}(1,1)
                \put(0.5,0.5){#2}
        \end{picture}
        &
        \begin{picture}(1,1)
                \put(0.5,0.5){#3}
        \end{picture}
        &
        \begin{picture}(1,1)
                \put(0.5,0.5){#4}
        \end{picture}
}
\caption{Particle States for LB1 (and LB2).}
\label{LB1:states}
\begin{center}
\begin{tabular}{||c|c|c|c||} \hline\hline
\multicolumn{4}{||c||}{Particle State $p$} \\ \hline
0 & 1 & 2 & 3 \\ \hline\hline
\Entry  {\PICSINN{\line}{\vector}{\line}{\line}{\Widths}}
        {\PICSINN{\vector}{\line}{\line}{\line}{\Widths}}
        {\PICSINN{\line}{\line}{\line}{\vector}{\Widths}}
        {\PICSINN{\line}{\line}{\vector}{\line}{\Widths}}
\\ \hline\hline
\end{tabular}
\end{center}
\end{table}
\begin{table}[p]
\newcommand{\PICINN}[5]{\setlength{\unitlength}{#5}
  \begin{picture}(0,0)\thicklines
    \put( 0.0,-0.5){#1(0,1){0.5}}
    \put(-0.5, 0.0){#2(1,0){0.5}}
    \put( 0.0, 0.5){#3(0,-1){0.5}}
    \put( 0.5, 0.0){#4(-1,0){0.5}}
  \end{picture}
}
\newcommand{\PICOUT}[5]{\setlength{\unitlength}{#5}
  \begin{picture}(0,0)\thicklines
    \put(0,0){#1(0,-1){0.5}}
    \put(0,0){#2(-1,0){0.5}}
    \put(0,0){#3(0,1){0.5}}
    \put(0,0){#4(1,0){0.5}}
  \end{picture}
}
\thinlines
\setlength{\unitlength}{4em}
\newlength{\Width}
\setlength{\Width}{0.8\unitlength}
\def\Entry#1#2#3{
        \begin{picture}(0,1)
                \put(0,0.5){\makebox(0,0){{\shortstack{#1}}}}
        \end{picture}
        &
        \begin{picture}(1,1)
                \put(0.5,0.5){#2}
        \end{picture}
        &
        \begin{picture}(1,1)
                \put(0.5,0.5){#3}
        \end{picture}
}
\renewcommand{\thefootnote}{\fnsymbol{footnote}}
\caption{Collision Rules for LB1.  Configurations that involve
changing particles' directions are marked ``\protect\footnotemark[1]''.}
\label{rules2}
\begin{center}
\begin{tabular}{||c||c|c||} \hline\hline
{~~Configuration~~}&{Pre-Collision State}&{Post-Collision State}
\\ \hline\hline
\Entry  {No particles}
        {\PICINN{\line}{\line}{\line}{\line}{\Width}}
        {\PICOUT{\line}{\line}{\line}{\line}{\Width}}
\\ \hline
\Entry  {One particle}
        {\PICINN{\line}{\vector}{\line}{\line}{\Width}}
        {\PICOUT{\line}{\line}{\line}{\vector}{\Width}}
\\ \hline
\Entry  {Two orthogonal \\ particles\footnotemark[1]}
        {\PICINN{\line}{\vector}{\vector}{\line}{\Width}}
        {\PICOUT{\line}{\vector}{\vector}{\line}{\Width}}
\\ \hline
\Entry  {Two head-on \\ particles\footnotemark[1]}
        {\PICINN{\line}{\vector}{\line}{\vector}{\Width}}
        {\PICOUT{\vector}{\line}{\vector}{\line}{\Width}}
\\ \hline
\Entry  {Three particles}
        {\PICINN{\vector}{\vector}{\vector}{\line}{\Width}}
        {\PICOUT{\vector}{\line}{\vector}{\vector}{\Width}}
\\ \hline
\Entry  {Four particles}
        {\PICINN{\vector}{\vector}{\vector}{\vector}{\Width}}
        {\PICOUT{\vector}{\vector}{\vector}{\vector}{\Width}}
\\ \hline\hline
\end{tabular}
\end{center}
\renewcommand{\thefootnote}{\arabic{footnote}}
\end{table}
\begin{table}
\caption{Collision Rules for LB1.}
\label{rules}
\begin{center}
\begin{tabular}
      {||r@{\hspace{1.25em}}||c@{\hspace{1.25ex}}c@{\hspace{1.25ex}}
         c@{\hspace{1.25ex}}c|c@{\hspace{1.25ex}}c@{\hspace{1.25ex}}
         c@{\hspace{1.25ex}}c|c||} \hline\hline
\multicolumn{1}{||c||}{Rule} &
\multicolumn{4}{c|}{$n$} &
\multicolumn{4}{c|}{${n'}$} &
$\SS(n,n')$ \\ \cline{2-9}
\multicolumn{1}{||c||}{\mbox{}}
       & $n(0)$  & $n(1)$  & $n(2)$  & $n(3)$  &
         $n'(0)$ & $n'(1)$ & $n'(2)$ & $n'(3)$ & \mbox{}
\\ \hline \hline
 0 & 0&0&0&0 & 0&0&0&0 & 1 \\
 1 & 0&0&0&1 & 0&0&0&1 & 1 \\
 2 & 0&0&1&0 & 0&0&1&0 & 1 \\
 3 & 0&0&1&1 & 1&1&0&0 & 1 \\
 4 & 0&1&0&0 & 0&1&0&0 & 1 \\
 5 & 0&1&0&1 & 1&0&1&0 & 1 \\
 6 & 0&1&1&0 & 1&0&0&1 & 1 \\
 7 & 0&1&1&1 & 0&1&1&1 & 1 \\
 8 & 1&0&0&0 & 1&0&0&0 & 1 \\
 9 & 1&0&0&1 & 0&1&1&0 & 1 \\
10 & 1&0&1&0 & 0&1&0&1 & 1 \\
11 & 1&0&1&1 & 1&0&1&1 & 1 \\
12 & 1&1&0&0 & 0&0&1&1 & 1 \\
13 & 1&1&0&1 & 1&1&0&1 & 1 \\
14 & 1&1&1&0 & 1&1&1&0 & 1 \\
15 & 1&1&1&1 & 1&1&1&1 & 1 \\ \hline\hline
\end{tabular}
\end{center}
\end{table}
Here,
\begin{equation}
  \label{velocity}
        \v(0) = \left [ \begin{array}{r}  1 \\  0 \end{array} \right ] \,, ~~
        \v(1) = \left [ \begin{array}{r}  0 \\  1 \end{array} \right ] \,, ~~
        \v(2) = \left [ \begin{array}{r} -1 \\  0 \end{array} \right ] \,, ~~
        \v(3) = \left [ \begin{array}{r}  0 \\ -1 \end{array} \right ] \,.
\end{equation}
The collision rules are illustrated in Table \ref{rules2} and
are formally tabulated in Table \ref{rules}.
Detailed balance is verified by examining each collision rule.  For
example, comparing rule 3 with rule 12 yields
$$
  \SS((0,0,1,1),(1,1,0,0)) = \SS((1,1,0,0),(0,0,1,1)) \,.
$$
Semidetailed balance is verified in the same manner.
If $F(k)= F_k, k=0,1,2,3$, then
the generalized Boltzmann collision operator $\CC$ is given by
\begin{eqnarray}
  \label{coll-lb1}
  &&
\\
  \nonumber
  \CC(F)(k)
  & = & \FBar_k \FBar_{k+1} F_{k+2} F_{k+3}
        + \FBar_k F_{k+1} \FBar_{k+2} F_{k+3}
\\
  \nonumber
  &   & + \FBar_k F_{k+1} F_{k+2} \FBar_{k+3}
        - F_k \FBar_{k+1} \FBar_{k+2} F_{k+3}
\\
  \nonumber
  &   & - F_k \FBar_{k+1} F_{k+2} \FBar_{k+3}
        - F_k F_{k+1} \FBar_{k+2} \FBar_{k+3}
\end{eqnarray}
where the sub-indices are evaluated modulo $4$. Note that $\CC$
conserves mass.

We know from the $H$-theorem (Theorem \ref{H-thm} (i)) that at a local
equilibrium, $\CC(F)=0\,$.  Thus, at a local equilibrium we have
\begin{eqnarray*}
  \CC(F)(0) = \CC(F)(1) & \Rightarrow & F_0 = F_1 \,,
\\
  \CC(F)(0) = \CC(F)(2) & \Rightarrow & F_0 = F_2 \,,
\\
  \CC(F)(0) = \CC(F)(3) & \Rightarrow & F_0 = F_3 \,.
\end{eqnarray*}
That is,
\begin{equation}
  \label{equil2}
  F_0 = F_1 = F_2 = F_3 \equiv u \,.
\end{equation}
If we take partial derivatives of (\ref{coll-lb1}) and evaluate them
at a local equilibrium (noting that (\ref{equil2}) holds), then the
linearized collision operator is the singular symmetric matrix
$$
 \LL_M \equiv \LL = - \frac{\nu}{4}
         \left[ \begin{array}{rrrr}
                        -3  &  1  &  1  &  1  \\
                         1  & -3  &  1  &  1  \\
                         1  &  1  & -3  &  1  \\
                         1  &  1  &  1  & -3  \end{array} \right] \,,
$$
where $\nu=-4u(1-u)$.  The eigenvalues of $\LL$ are
$$
  (\lambda_0,\lambda_1,\lambda_2,\lambda_3) = (0,\nu,\nu,\nu)
$$
and the associated unnormalized eigenmatrix is
\begin{equation}
  \label{eigen}
  {\bf Q}
  = \left[ \q_0 \,,~\q_1 \,,~\q_2 \,,~\q_3 \right]
  = \left[ \begin{array}{rrrr}
                    1  &  1  &  0  &  1  \\
                    1  &  0  &  1  & -1  \\
                    1  & -1  &  0  &  1  \\
                    1  &  0  & -1  & -1  \end{array} \right] \,.
\end{equation}
By Theorem \ref{thm3.3}, ${\bf E} = \NULL(\LL) = \SPAN[{\bf q}_0]$.
The pseudoinverse of $\LL$ is given by
\begin{equation}
\label{pseudo-lb1}
  \LL^{-1} = \nu^{-1}
             \left[ \tfrac12 \left( \q_1 \q_1^t + \q_2 \q_2^t \right)
                    + \tfrac14 \q_3 \q_3^t \right] \,.
\end{equation}
Clearly, the lattice gas is diffusive (Definition \ref{diffusive}).
A Hilbert expansion (\ref{series}) for LB1 is determined where we
take $\CC^{(0)}=\CC$ and $\CC^{(1)}$ to be the null operator.
It follows from (\ref{h0}) that
\begin{equation}
\label{h0-lb1}
  h^{(0)} = u(t,\x) \q_0 \,.
\end{equation}
{}From (\ref{h1}), (\ref{pseudo-lb1}), and (\ref{h0-lb1}),
\begin{equation}
\label{h1-lb1}
h^{(1)} = L \nu^{-1} [ u_x \q_1 + u_y \q_2] + \beta^{(1)} \q_0
\end{equation}
for some $\beta^{(1)} \in \R$.

Equation (\ref{order2}) yields
$$
\LL h^{(2)} = a^{(2)}-r^{(2)}
$$
where
\begin{eqnarray*}
a^{(2)} &=& [T\partial_t + \tfrac12 L^2 (\v \cdot \nabla)^2] h^{(0)} +
L(\v \cdot \nabla) h^{(1)} \\
\\
r^{(2)} &=&- \tfrac12 \DD^2 \CC(h^{(0)}) \cdot h^{(1)} h^{(1)}.
\end{eqnarray*}
If we substitute (\ref{h0-lb1}) and (\ref{h1-lb1}) into the
solvability condition (\ref{solve2}), we get the nonlinear diffusion
equation
\begin{equation}
  \label{eqn1}
  \del_t u = \mu \nabla \cdot D(u) \GRAD u \,,~~
  D(u) = - \left( \frac{1}{\nu} + \frac{1}{2} \right), ~
   \mu = \frac{L^2}{2T} \,.
\end{equation}

The solvability condition for $h^{(3)}$ yields
\begin{equation}
\label{b1-lba}
\del_t\beta^{(1)} = \mu \nabla \cdot \left [ D(u) \nabla \beta^{(1)}
+D'(u) \beta^{(1)} \nabla u \right ]
\end{equation}
so that we can set $\beta^{(1)} = 0$ in (\ref{h1-lb1}).

The solvability condition for $h^{(4)}$ is
\begin{equation}
  \label{eqn2}
  \del_t \beta^{(2)}
   =  \mu \DIV \left[ D(u) \GRAD \beta^{(2)}
                        + D'(u) \beta^{(2)} \GRAD u \right]
        - {\cal F} \,.
\end{equation}
where ${\cal F}$ is a smooth function of $u$ and its derivatives, and
the solvability condition for $h^{(5)}$ is
\begin{equation}
\label{b3-lb1}
\del_t \beta^{(3)} = \mu \DIV \left [ D(u) \GRAD \beta^{(3)}
+ D'(u) \beta^{(3)} \GRAD u \right ].
\end{equation}
We can thus take $\beta^{(3)} = 0$.
The expressions for $h^{(j)}$, $j = 0,1,2,3$ are
given in Appendix A.

  We now consider the grid function
$$
  H^{(3)}(m,{\bf i},\cdot)
  = \sum_{k=0}^3 h^{(k)}(m,{\bf i},\cdot) \delta^k \,.
$$
Then,
$$
  \AA H^{(3)} - H^{(3)} - \CC(H^{(3)})
  = -{\bf T}(H^{(3)}) + {\rm higher~order~terms} \,,
$$
where
$$
  {\bf T}(H^{(3)}) = \sum_{j=0}^4 {\bf T}^{(j)} \delta^j \,.
$$
Since ${\bf T}^{(0)} = \LL h^{(0)}$ and
${\bf T}^{(j)}=\LL h^{(j)} - a^{(j)} + s^{(j)} \,$ $j=1,2,3,$ it follows that
${\bf T}^{(j)}=0$, $j=0,1,2,3$.  Moreover,
$$
  {\bf T}^{(4)} = - \sum_{k=0}^3 \tilde{c}_k^{(4)} \q_k
                = - \tilde{c}_3^{(4)} \q_3 \,,
$$
where $\tilde{c}_3^{(4)}$ is given in Appendix C. Since
$u\in C^4([0,1]^2,(0,1))$ and $\beta^{(2)} \in C^2(\R^2,\R)$,
we have
$$
  \TT(H^{(3)}) = {\cal O}(\delta^4) \,,
$$
and the method is consistent.

  Stability of LB1 will follow from Theorem \ref{thm53} once we have
established an invariant domain of monotonicity for the method.

\begin{theorem} \label{thm6.1}
The set $\EE = [M_-,M_+]^{4}$, where
$$
  M_+  =  \tfrac12 (1+1/\sqrt{5})  \qquad \mbox{and} \qquad
  M_-  =  \tfrac12 (1-1/\sqrt{5}) \,,
$$
is a domain of monotonicity for LB1.  Further, the collision
operator is zero at the extreme points of $\EE$.
\end{theorem}

\begin{proof}
The elements of the Jacobian matrix $\JJ[\BB]$ are
(using (\ref{coll-lb1}))
\[
\begin{array}{lcl}
  \JJ[\BB]_{k,k}
  & = & 1 + F_{k+2} F_{k+3} + F_{k+1} F_{k+3}
          + F_{k+1} F_{k+2} - F_{k+1} - F_{k+2} - F_{k+3} \,,
\\
  \JJ[\BB]_{k,k+1}
  & = & -3 F_{k+2} F_{k+3} + F_k F_{k+3} + F_k F_{k+2}
         + F_{k+2} + F_{k+3} - F_k \,,
\\
  \JJ[\BB]_{k,k+2}
  & = & -3 F_{k+1} F_{k+3} + F_k F_{k+3} + F_k F_{k+1}
         + F_{k+1} + F_{k+3} - F_k \,,
\\
  \JJ[\BB]_{k,k+3}
  & = & -3 F_{k+1} F_{k+2} + F_k F_{k+2} + F_k F_{k+1}
         + F_{k+1} + F_{k+2} - F_k \,,
\end{array}
\]
where the sub-indices are evaluated modulo $4$.
Note that $\JJ[\BB](F)$ will be nonnegative on $\EE$
if each of the functions
\begin{eqnarray*}
  f(x,y,z) & = & 1 + y z + x z + x y - x - y - z \,,
\\
  g(x,y,z) & = & -3 y z + x z + x y + z + y - x
\end{eqnarray*}
are nonnegative for $M_- \leq x,y,z \leq M_+$. This will be case if
the local and boundary minima of $f$ and $g$ are
nonnegative.

If we examine the gradients of these functions, then we
see that they are both zero at the point $x=y=z=\tfrac12$.  However,
this point is not a local extremum for either $f$ or $g$ since
their Hessian matrices $D^2f$, $D^2g$ are neither positive nor negative
definite there.
On the boundaries we have
$$
  0 \leq f,g \leq \tfrac25 \,.
$$
Hence, $\JJ[\BB](F) \geq 0$ for $F \in \EE$.

  Since the extreme points of $\EE$ are
$\MM_+ = M_+ \q_0$ and $\MM_- = M_- \q_0$, it
follows from (\ref{h0-lb1}) that
$$
  \CC(\MM_+) = \CC(\MM_-) = {\bf 0} \,,
$$
and the proof is complete.
\end{proof}

  We can now apply Theorem \ref{thm51} to show that the solution to
LB1 converges to the solution $u$ of (\ref{eqn1}) as the lattice
spacing is refined.

\subsubsection*{Numerical Verification}

  We compute solutions to the nonlinear diffusion equation
(\ref{eqn1}) for $(x,y)\in[0,1]\times[0,1]$, $t>0$,
$\mu=\tfrac{1}{2}$, periodic boundary conditions, and initial
condition
$$
  u(x,y,0) = \tfrac{1}{\sqrt{2}} \sin (2\pi x) \sin (2\pi y)
             + \tfrac12 \,.
$$
The solutions are computed using both LB1 and an explicit finite
difference method that is first order accurate in time and second
order accurate in space.  Note that the initial condition for the
lattice Boltzmann method lies within the domain of monotonicity of
Theorem \ref{thm6.1}.

  The lattice Boltzmann approximations on $N\times N$ grids ($N<256$)
at lattice point {\bf i} and time $m$ are computed by
$$
  (u^{(N)})_{{\bf i}}^m = \tfrac14 \< F \>\,,
$$
The finite difference solutions are computed only on a $256 \times
256$ grid and rendered on coarser grids by pointwise projection to
yield the solutions $(v^{(N)})_{{\bf i}}^m$.  Note that the finite
difference computed solution retains its accuracy when projected onto
the coarser grids.

\begin{figure}
\setlength{\Width}{0.475\textwidth}
\setlength{\unitlength}{\Width}
\def\Psfig#1{%
%
{\psfig{figure=psfigs/#1,width=\Width,bbllx=1.5in,bblly=0in,bburx=7in,bbury=5in,clip=}}}
\def\Put#1#2{\begin{picture}(0,0)
                   \put(0,#1){\makebox(0,0){#2}}
              \end{picture}}
\begin{center}
\begin{tabular}{||@{}c@{}|@{}c@{}||} \hline\hline
\setlength{\unitlength}{\Width}
\Psfig{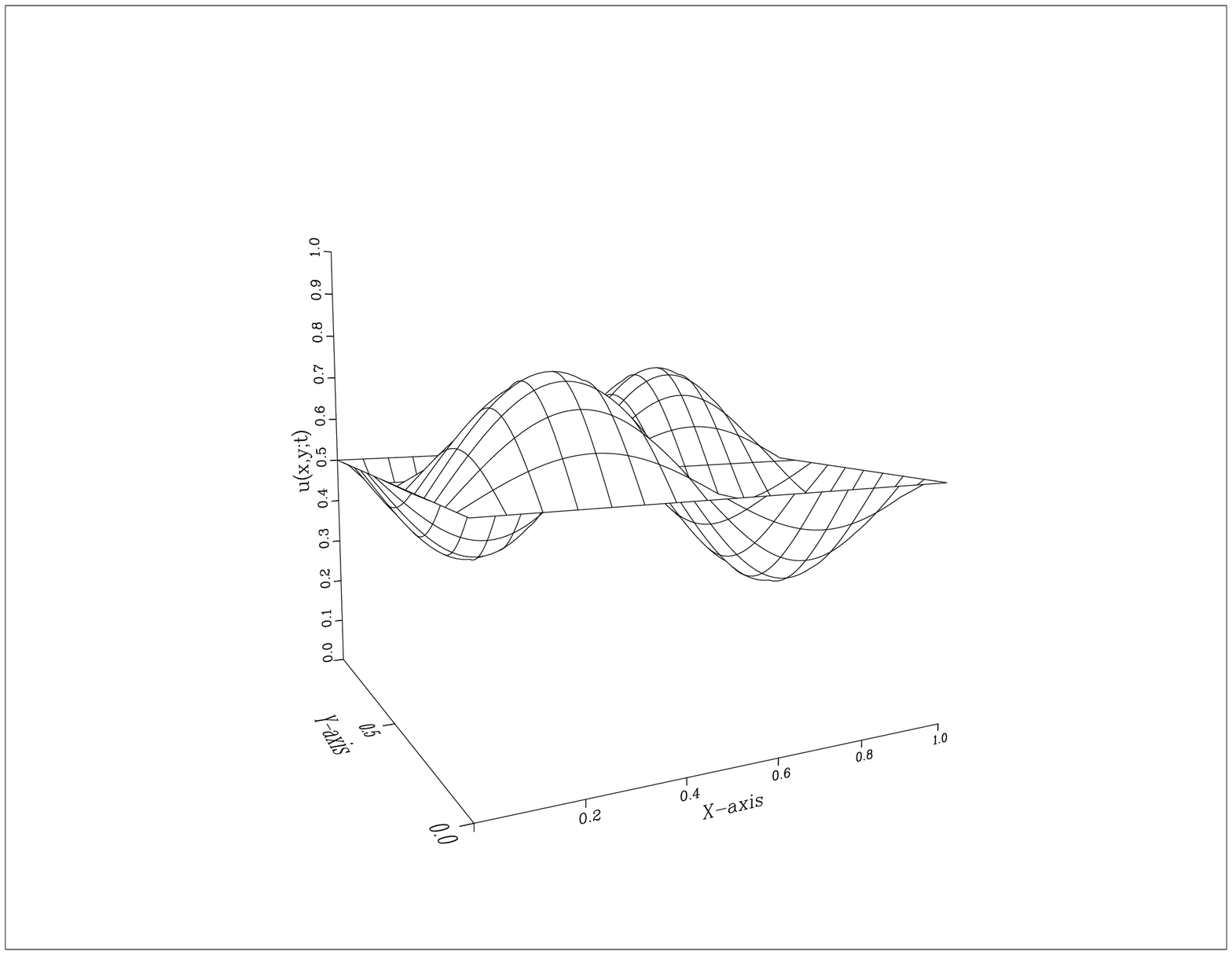} & \Psfig{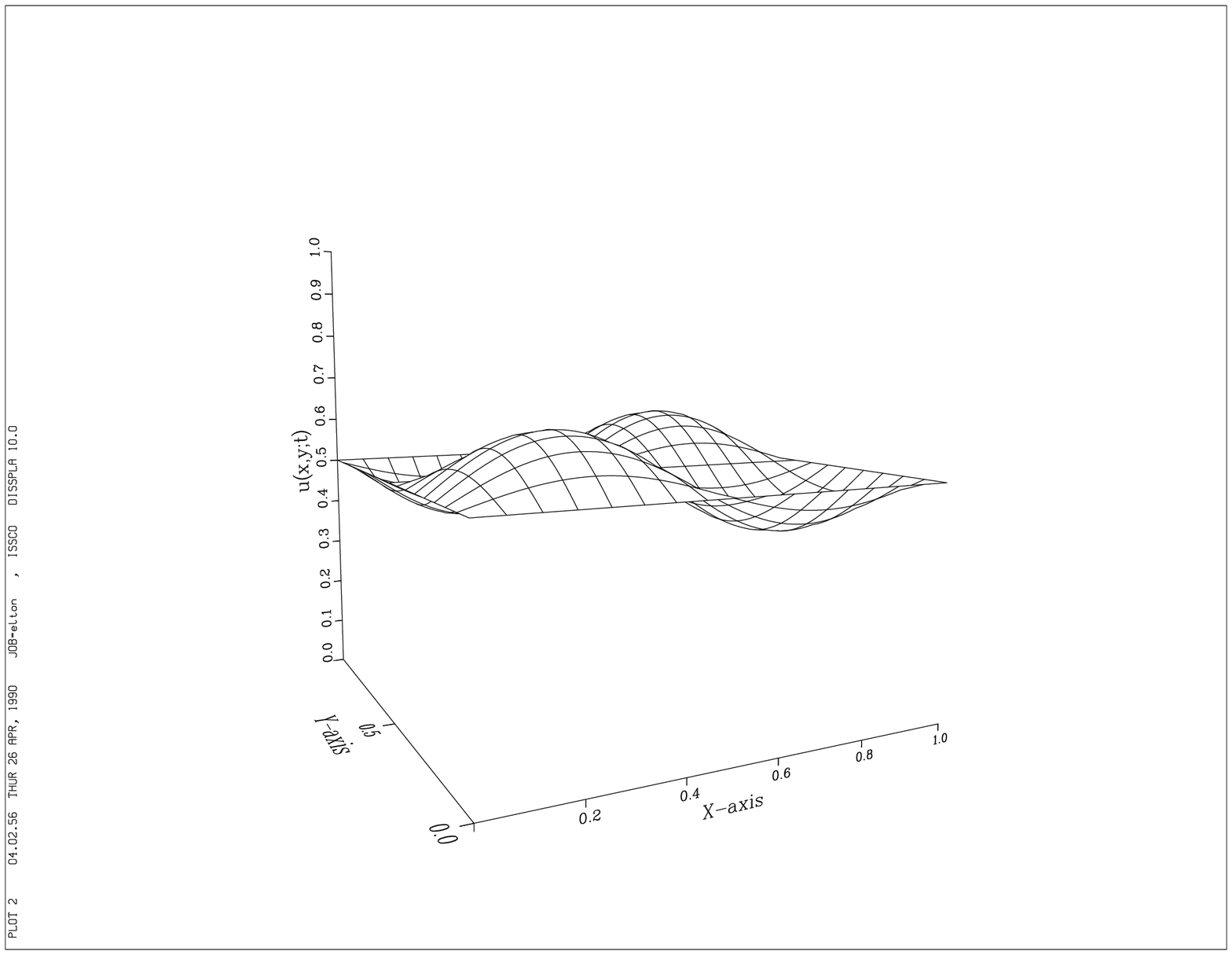} \\*[\baselineskip]
\Put{+0.05}{(a) $t=0$} &
\Put{+0.05}{(b) $t=1/32$} \\ \hline\hline
\end{tabular}
\end{center}
\caption{LB1:~~$u(t,x,y)$.}
\label{eg1}
\end{figure}

  Figure \ref{eg1} exhibits the solution for LB1 at times $t=0$ and
$t=1/32$.  Table \ref{tab2} and Figure \ref{fig8} support the
theoretical second-order accuracy where $E^{(N)}$ denotes the error
vector which has components
$(v^{(N)})_{{\bf i}}^m - (u^{(N)})_{{\bf i}}^m$.
Moreover, the ratio of the errors over grid sizes a factor of two
apart yields the expected ratio of four.
\begin{table}
\def\FixUpRatio#1{\raisebox{1ex}[4ex]{#1}}
\def\norm#1{\left\|{#1}\right\|}
\def\Norm#1#2{\norm{#1}_{#2}}
\def\lonenorm#1{\Norm{#1}{\ell_1}}
\def\linfnorm#1{\Norm{#1}{\ell_\infty}}
\caption{LB1:~~Norm comparisons }
\label{tab2}
\begin{center}
\renewcommand{\arraystretch}{1.25}
\begin{tabular}{||r||l|l|l|l||} \hline\hline
\mbox{} &
\multicolumn{4}{|c||}{\FixUpRatio{$\lonenorm{E^{(N)}}$}} \\ \cline{2-5}
\multicolumn{1}{||c||}{$N$} &
      $t=1/32$   & $t=1/16$  & $t=3/32$   & $t=1/8$    \\ \hline\hline
8   & 0.00241    & 0.00301   & 0.00241    & 0.00170    \\ \hline
16  & 0.000605   & 0.000759  & 0.000624   & 0.000449   \\ \hline
32  & 0.000151   & 0.000190  & 0.000157   & 0.000114   \\ \hline
64  & 0.0000366  & 0.0000465 & 0.0000387  & 0.0000279  \\ \hline
128 & 0.00000804 & 0.0000107 & 0.00000902 & 0.00000662 \\ \hline
\multicolumn{5}{c}{\mbox{}\vspace{-2ex}} \\ \hline
\multicolumn{1}{||c||}{$N$} &
\multicolumn{4}{|c||}{\FixUpRatio{$\lonenorm{E^{(N)}}/\lonenorm{E^{(2N)}}$}}
\\ \hline
8   & 3.991 & 3.961 & 3.867 & 3.776 \\ \hline
16  & 4.010 & 4.002 & 3.987 & 3.955 \\ \hline
32  & 4.124 & 4.078 & 4.051 & 4.062 \\ \hline
64  & 4.550 & 4.336 & 4.287 & 4.223 \\ \hline\hline
\end{tabular}
\end{center}
\end{table}
\begin{figure}
\def\norm#1{\left\|{#1}\right\|}
\def\Norm#1#2{\norm{#1}_{#2}}
\def\lonenorm#1{\Norm{#1}{\ell_1}}
\def\linfnorm#1{\Norm{#1}{\ell_\infty}}
\def\sfbtenpt{\sf}
\setlength{\Width}{0.45\textwidth}
\newcommand{\PSFIG}[1]{%
%
{\psfig{figure=psfigs/#1,bbllx=72bp,bblly=72bp,bburx=540bp,bbury=605bp,width=\Width,rheight=0.85\Width,clip=}}%
}
\newcommand{\LabelQ}[1]{%
\makebox[0in]{
\setlength{\unitlength}{\Width}
\begin{picture}(0,0)
  \put(+0.09  ,+0.7  ){\makebox(0,0){\sfbtenpt{\shortstack[c]{#1}}}}
  \put(+0.3   ,+0.05 ){\makebox(0,0){\sfbtenpt{$N$}}}
  \put(+0.95  ,+0.225){\makebox(0,0){\sfbtenpt{$t$}}}
\end{picture}}
}
\newcommand{\LabelR}[1]{%
\makebox[0in]{
\setlength{\unitlength}{\Width}
\begin{picture}(0,0)
  \put(+0.09  ,+0.725){\makebox(0,0){\sfbtenpt{\shortstack[c]{#1}}}}
  \put(+0.3   ,+0.05 ){\makebox(0,0){\sfbtenpt{$N\!:2N$}}}
  \put(+0.95  ,+0.225){\makebox(0,0){\sfbtenpt{$t$}}}
\end{picture}}
}
\begin{center}
\begin{tabular}{||c|c||} \hline\hline
\LabelQ{$\lonenorm{E^{(N)}}$}\PSFIG{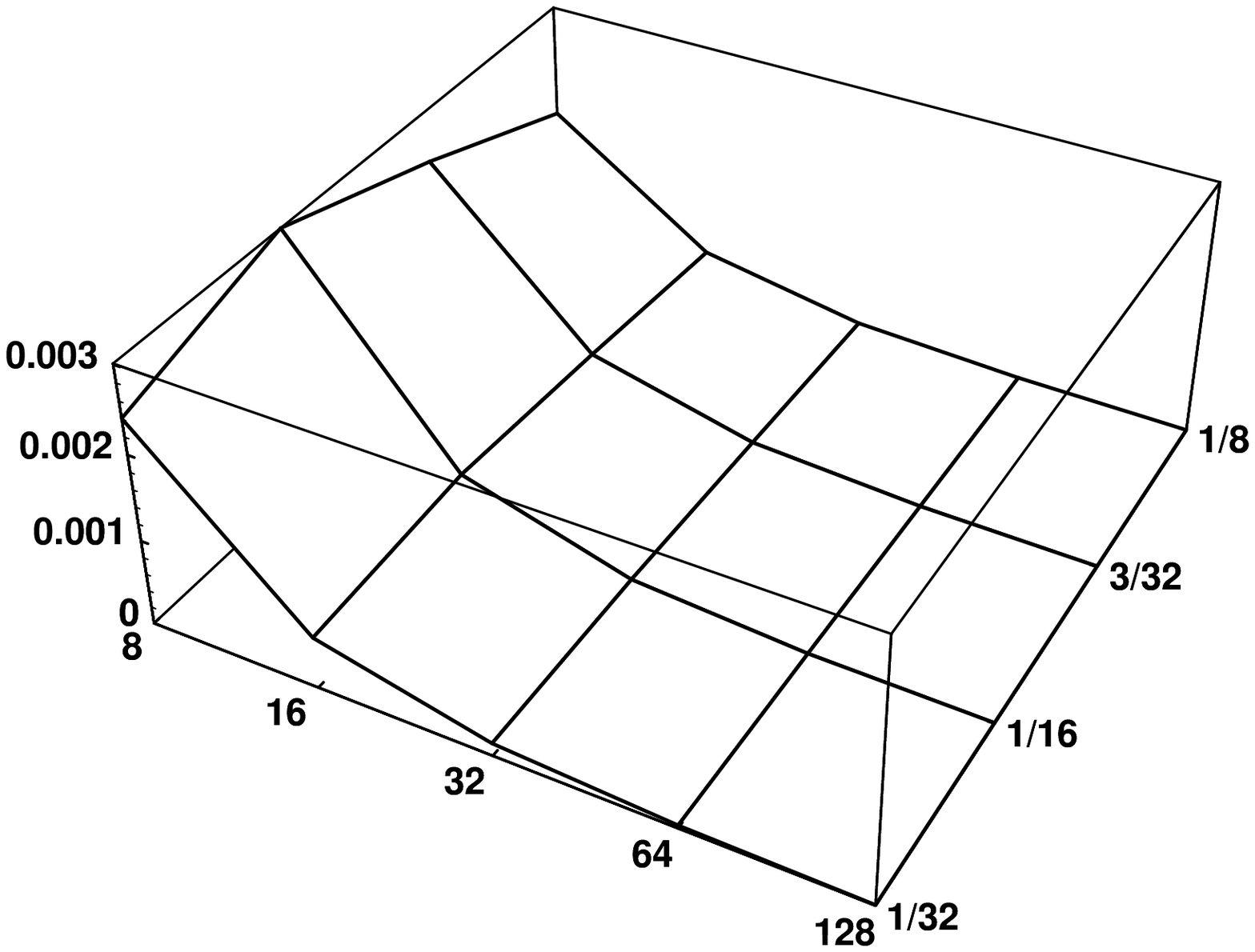} &
\LabelR{$\frac{\lonenorm{E^{(N)}}}{\lonenorm{E^{(2N)}}}$}\PSFIG{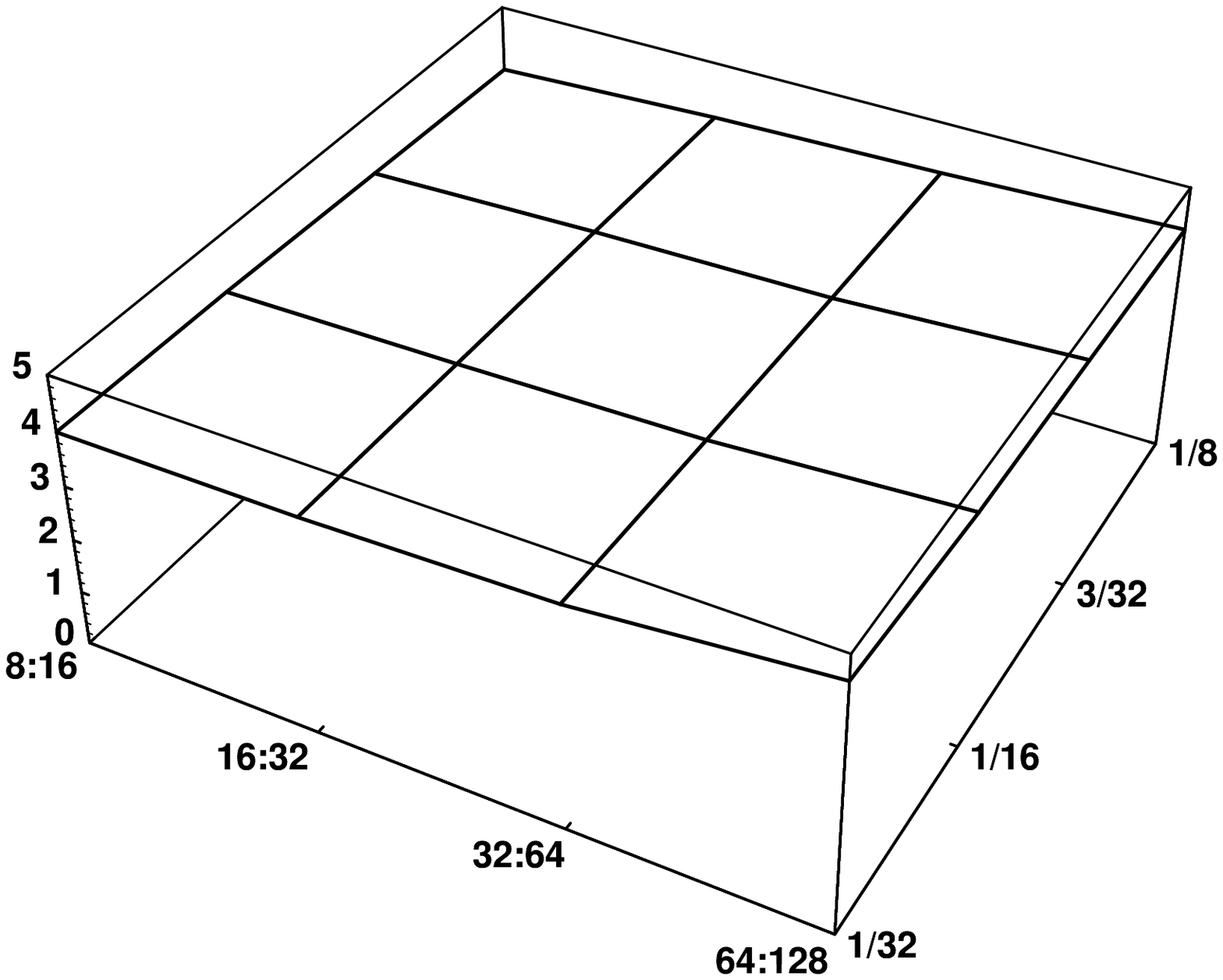}
\\
(a) & (b) \\ \hline\hline
\end{tabular}
\end{center}
\caption{%
LB1:~~Norm comparisons ,
(a) Errors; (b) error ratios.
}
\label{fig8}
\end{figure}


\clearpage
\section{Example - Another Nonlinear Diffusive System}

  We now consider a variation of LB1, called LB2, where again the
lattice gas is defined on a periodic square lattice, the possible
particle states are in Table \ref{LB1:states} and the velocity vectors
are given by (\ref{velocity}).  The collision rules for LB2 are given
in Table \ref{rules3}.

\begin{table}
\newcommand{\PICINN}[5]{\setlength{\unitlength}{#5}
  \begin{picture}(0,0)\thicklines
    \put( 0.0,-0.5){#1(0,1){0.5}}
    \put(-0.5, 0.0){#2(1,0){0.5}}
    \put( 0.0, 0.5){#3(0,-1){0.5}}
    \put( 0.5, 0.0){#4(-1,0){0.5}}
  \end{picture}
}
\newcommand{\PICOUT}[5]{\setlength{\unitlength}{#5}
  \begin{picture}(0,0)\thicklines
    \put(0,0){#1(0,-1){0.5}}
    \put(0,0){#2(-1,0){0.5}}
    \put(0,0){#3(0,1){0.5}}
    \put(0,0){#4(1,0){0.5}}
  \end{picture}
}
\thinlines
\setlength{\unitlength}{4em}
\setlength{\Width}{0.8\unitlength}
\def\Entry#1#2#3{
        \begin{picture}(0,1)
                \put(0,0.5){\makebox(0,0){{\shortstack{#1}}}}
        \end{picture}
        &
        \begin{picture}(1,1)
                \put(0.5,0.5){#2}
        \end{picture}
        &
        \begin{picture}(1,1)
                \put(0.5,0.5){#3}
        \end{picture}
}
\renewcommand{\thefootnote}{\fnsymbol{footnote}}
\caption{Collision Rules for LB2.  Configurations that involve changing
particles' directions are marked ``\protect\footnotemark[1]''.}
\label{rules3}
\begin{center}
\begin{tabular}{||c||c|c||} \hline\hline
{~~Configuration~~}&{Pre-Collision State}&{Post-Collision State} \\
\hline\hline
\Entry  {No particles}
        {\PICINN{\line}{\line}{\line}{\line}{\Width}}
        {\PICOUT{\line}{\line}{\line}{\line}{\Width}}
\\ \hline
\Entry  {One particle\footnotemark[1]}
        {\PICINN{\line}{\vector}{\line}{\line}{\Width}}
        {\PICOUT{\line}{\vector}{\line}{\line}{\Width}}
\\ \hline
\Entry  {Two orthogonal \\ particles\footnotemark[1]}
        {\PICINN{\line}{\vector}{\vector}{\line}{\Width}}
        {\PICOUT{\line}{\vector}{\vector}{\line}{\Width}}
\\ \hline
\Entry  {Two head-on \\ particles\footnotemark[1]}
        {\PICINN{\line}{\vector}{\line}{\vector}{\Width}}
        {\PICOUT{\vector}{\line}{\vector}{\line}{\Width}}
\\ \hline
\Entry  {Three particles}
        {\PICINN{\vector}{\vector}{\vector}{\line}{\Width}}
        {\PICOUT{\vector}{\line}{\vector}{\vector}{\Width}}
\\ \hline
\Entry  {Four particles}
        {\PICINN{\vector}{\vector}{\vector}{\vector}{\Width}}
        {\PICOUT{\vector}{\vector}{\vector}{\vector}{\Width}}
\\ \hline\hline
\end{tabular}
\end{center}
\renewcommand{\thefootnote}{\arabic{footnote}}
\end{table}
The collision operator for LB2 is in semidetailed balance and is given by
\begin{equation}
\label{col-lb2}
\begin{array} {ccc}
  \CC(F)(k) & = & \bar{F}_k \bar{F}_{k+1}
                    F_{k+2} F_{k+3}
                    + \bar{F}_k F_{k+1}
                      \bar{F}_{k+2} F_{k+3} \\
              &   & + \bar{F}_k F_{k+1}
                      F_{k+2} \bar{F}_{k+3}
                    - F_k \bar{F}_{k+1}
                      \bar{F}_{k+2} F_{k+3} \\
              &   & - F_k \bar{F}_{k+1}
                      F_{k+2} \bar{F}_{k+3}
                    - F_k F_{k+1}
                      \bar{F}_{k+2} \bar{F}_{k+3} \\
              &   & + F_k F_{k+1}
                      \bar{F}_{k+2} F_{k+3}
                    - F_k \bar{F}_{k+1}
                      \bar{F}_{k+2} \bar{F}_{k+3} \,,
\end{array}
\end{equation}
where the indices are evaluated modulo $4$ and, as before, $F_k=F(k)$.
Note that the collision operator for LB2 is identical to the collision
operator for LB1 with the exception of the last two terms.  The
lattice gas is diffusive and the collision operator conserves mass.

  At a local equilibrium of $\CC$ we have
$$
  F_0 = F_1 = F_2 = F_3 = u \,.
$$
The linearized collision operator is given by
$$
  \LL = - (1-u) \left[ \begin{array}{rrrr}
                       -(2u-1) &    u    &    1    &    u      \\
                           u   & -(2u-1) &    u    &    1      \\
                           1   &    u    & -(2u+1) &    1      \\
                           u   &    1    &    u    & -(2u+1) \end{array}
  \right] \,.
$$
The eigenvalues of $\LL$ are
$$
  (\lambda_0,\lambda_1,\lambda_2,\lambda_3)
  = (0,-2(1-u^2),-2(1-u^2),-4u(1-u)) \,
$$
and the unnormalized eigenmatrix is the same as for LB1, cf. (\ref{eigen}).
Let $\nu = \lambda_2 = -2(1-u^2)$.

  The determination of the Hilbert expansion for LB2 proceeds in the
same manner as for LB1.  In this case, we get
\begin{eqnarray}
  \nonumber
  h^{(0)} &=& u(t,\x) {\bf q}_0 \,,
\\
  \label{sol-lb2}
  u_t &=& \mu \GRAD \cdot D(u) \GRAD u \,,~~
  D(u) = -\left ( \frac{1}{\nu} + \frac{1}{2} \right )~,
  ~~\mu = \frac{L^2}{2T} \,,
\\
  \nonumber
  \beta^{(1)} &=& 0 \,,
\\
  \nonumber
  \partial_t \beta^{(2)}
  &=& \frac{L^2}{2T} \DIV \left[ D(u) \GRAD \beta^{(2)}
      + D'(u) \beta^{(2)} \GRAD u \right] - {\cal F} \,,
\\
  \nonumber
  \beta^{(3)} &=& 0 \,,
\end{eqnarray}
where ${\cal F}$ is a smooth function of $u$ and its derivatives.
Consistency of LB2 follows in the same manner as in LB1. Indeed,
$$
  \AA H^{(3)} - H^{(3)} - \CC(H^{(3)})
  = {\bf T}(H^{(3)}) = - \tilde{c}_3^{(4)}{\bf q}_3 \,,
$$
where $\tilde{c}_3^{(4)}$ is given in Appendix C.  Consistency then
follows from the fact that ${\bf T}(H^{(3)})={\cal O}(\delta^4)$.

\begin{theorem} The set $\EE=[M_-,M_+]^4$, where
$$
  M_+ = \tfrac{2}{3} \qquad \mbox{and} \qquad M_- = \tfrac{5}{6} \,,
$$
is an domain of monotonicity for LB2.  Further, the collision
operator is zero at the extreme points of $\EE$.
\end{theorem}
\begin{proof}
The elements of the Jacobian matrix $\JJ[\BB]$ are (using (\ref{col-lb2})
\[
\begin{array}{lcl}
  \JJ[\BB]_{k,k}
  &=& F_{k+2} F_{k+3} + F_{k+1} F_{k+2} - F_{k+2} \,,
\\
  \JJ[\BB]_{k,k+1}
  &=& -2 F_{k+2} F_{k+3} + F_k F_{k+2} + F_{k+3} \,,
\\
  \JJ[\BB]_{k,k+2}
  &=& -2 F_{k+1} F_{k+3} + F_k F_{k+3} + F_k F_{k+1} - F_k + 1 \,,
\\
  \JJ[\BB]_{k,k+3}
  &=& -2 F_{k+1} F_{k+2} + F_k F_{k+2} + F_{k+1} \,,
\end{array}
\]
where the sub-indices are evaluated modulo $4$.  We show that
$\JJ[\BB]$ is nonnegative on $\EE$ by showing that each of the
functions
\begin{eqnarray*}
  f(x,y,z) &=& yz +xz - y \\
  g(x,y,z) &=& -2yz + xy + z \\
  h(x,y,z) &=& -2yz + xz + xy - x + 1
\end{eqnarray*}
are nonnegative on $M_- \leq x,y,z \leq M_+$.

  The gradients of these functions are never zero in the domain of
restriction for the independent variables and so the extrema of these
functions is located on the boundaries.  A lengthy analysis of these
functions shows that on these boundaries we have
$$
  0 \leq f,g,h \leq \tfrac{11}{36},
$$
cf.\ \cite{Elton90}.  Hence, $\JJ[\BB](F)\geq 0$ for $F\in\EE$.

Since the extreme points of $\EE$ are $M_+{\bf q}_0$ and
$M_-{\bf Q}_0$, it follows $\CC(\MM_+)=\CC(\MM_-)=0$ and the proof is
complete.
\end{proof}

  We can now apply Theorem \ref{thm51} to show that the solution to
LB2 converges to the solution $u$ of (\ref{sol-lb2}) as the lattice
spacing is refined.

\subsubsection*{Numerical Verification}

\begin{figure}
\setlength{\Width}{0.475\textwidth}
\setlength{\unitlength}{\Width}
\def\Psfig#1{%
%
{\psfig{figure=psfigs/#1,width=\Width,bbllx=1.5in,bblly=0in,bburx=7in,bbury=5in,clip=}}
}
\def\Put#1#2{
        \begin{picture}(0,0)
                \put(0,#1){\makebox(0,0){#2}}
        \end{picture}
}
\begin{center}
\begin{tabular}{||@{}c@{}|@{}c@{}||} \hline\hline
\setlength{\unitlength}{\Width}
\Psfig{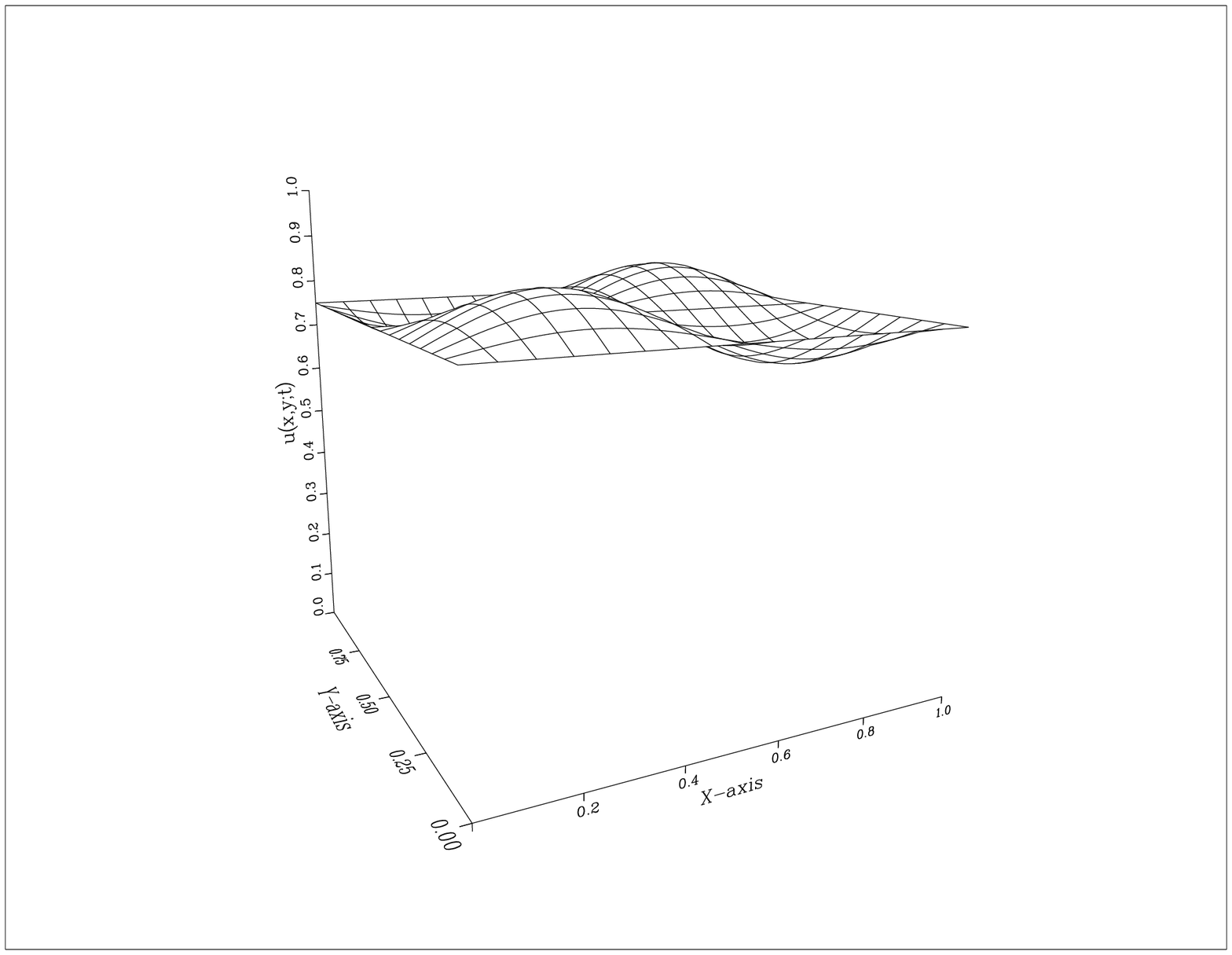} & \Psfig{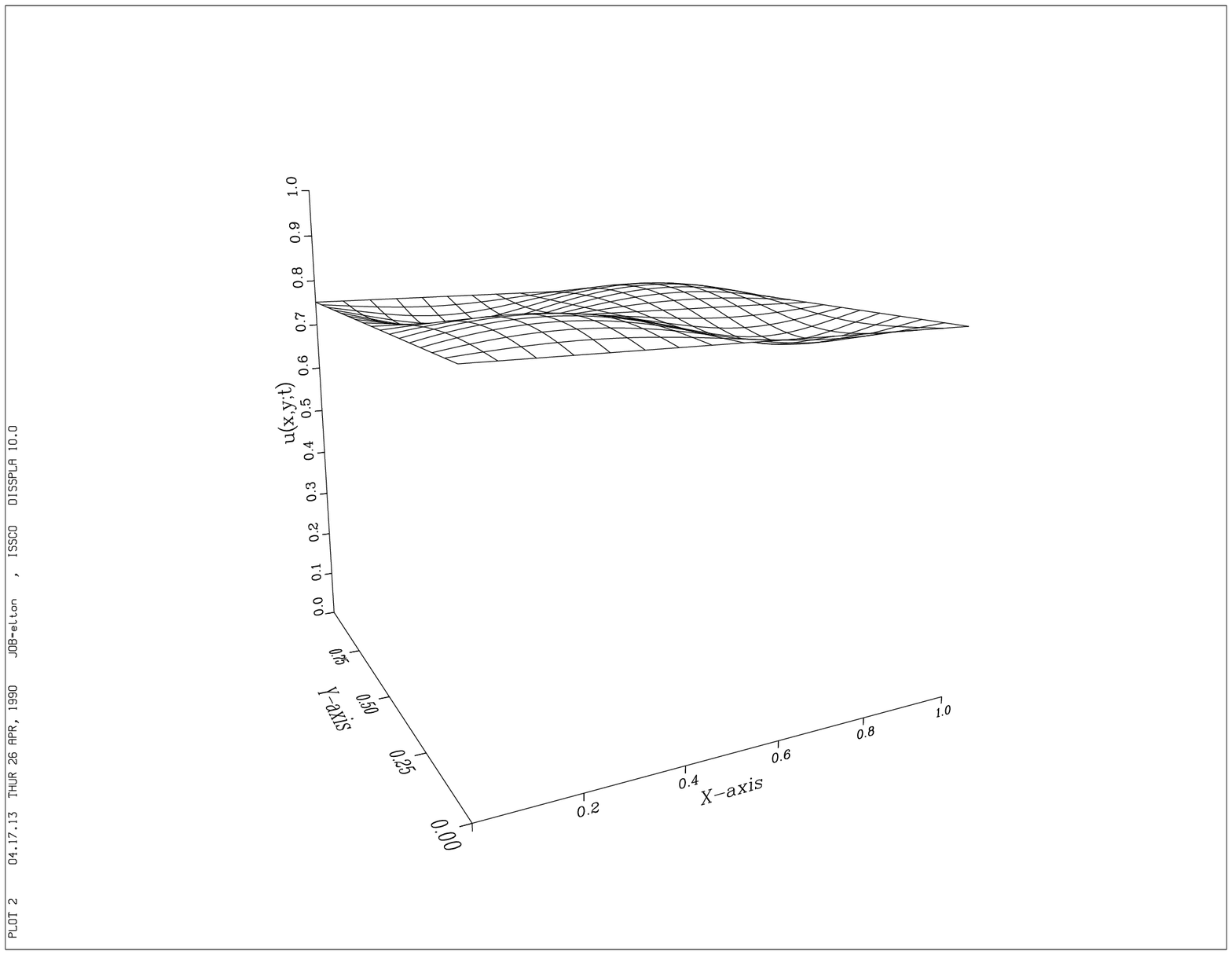} \\*[\baselineskip]
\Put{+0.05}{(a) $t=0$} &
\Put{+0.05}{(b) $t=1/32$} \\ \hline\hline
\end{tabular}
\end{center}
\caption{LB2:~~$u(t,x,y)$.}
\label{eg2}
\end{figure}

  We repeat the same experiments as was done for LB1.  That is, we
solve (\ref{sol-lb2}) with the same boundary conditions and the
initial condition
$$
  u(x,y,0) = \tfrac12 \sin(2\pi x)\sin(2\pi y) + \tfrac34 \,.
$$
The results are given in Figures \ref{eg2} and \ref{fig9} and in Table
\ref{tab4}.
\clearpage
\begin{table}
\def\FixUpRatio#1{\raisebox{1ex}[4ex]{#1}}
\def\norm#1{\left\|{#1}\right\|}
\def\Norm#1#2{\norm{#1}_{#2}}
\def\lonenorm#1{\Norm{#1}{\ell_1}}
\def\linfnorm#1{\Norm{#1}{\ell_\infty}}
\caption{LB2:~~Norm comparisons for $A=1/12$ and $B=3/4$.}
\label{tab4}
\begin{center}
\renewcommand{\arraystretch}{1.25}
\begin{tabular}{||r||l|l|l|l||} \hline\hline
\mbox{} &
\multicolumn{4}{|c||}{\FixUpRatio{$\lonenorm{E^{(N)}}$}} \\ \cline{2-5}
\multicolumn{1}{||c||}{$N$} &
      $t=1/32$   & $t=1/16$   & $t=3/32$   & $t=1/8$    \\ \hline\hline
8   & 0.000692   & 0.00202    & 0.00144    & 0.000827   \\ \hline
16  & 0.000276   & 0.000443   & 0.000321   & 0.000198   \\ \hline
32  & 0.0000722  & 0.000106   & 0.0000782  & 0.0000491  \\ \hline
64  & 0.0000180  & 0.0000264  & 0.0000197  & 0.0000123  \\ \hline
128 & 0.00000463 & 0.00000668 & 0.00000502 & 0.00000316 \\ \hline
\multicolumn{5}{c}{\mbox{}\vspace{-2ex}} \\ \hline
\multicolumn{1}{||c||}{$N$} &
\multicolumn{4}{|c||}{\FixUpRatio{$\lonenorm{E^{(N)}}/\lonenorm{E^{(2N)}}$}}
\\ \hline
8   & 2.504 & 4.572 & 4.480 & 4.186 \\ \hline
16  & 3.827 & 4.166 & 4.106 & 4.021 \\ \hline
32  & 4.009 & 4.028 & 3.971 & 3.995 \\ \hline
64  & 3.893 & 3.948 & 3.924 & 3.893 \\ \hline\hline
\end{tabular}
\end{center}
\end{table}
\begin{figure}
\def\norm#1{\left\|{#1}\right\|}
\def\Norm#1#2{\norm{#1}_{#2}}
\def\lonenorm#1{\Norm{#1}{\ell_1}}
\def\linfnorm#1{\Norm{#1}{\ell_\infty}}
\def\sfbtenpt{\sf}
\setlength{\Width}{0.45\textwidth}
\newcommand{\PSFIG}[1]{%
%
{\psfig{figure=psfigs/#1,bbllx=72bp,bblly=72bp,bburx=540bp,bbury=605bp,width=\Width,rheight=0.85\Width,clip=}}%
}
\newcommand{\LabelQ}[1]{%
\makebox[0in]{
\setlength{\unitlength}{\Width}
\begin{picture}(0,0)
  \put(+0.09  ,+0.7  ){\makebox(0,0){\sfbtenpt{\shortstack[c]{#1}}}}
  \put(+0.3   ,+0.05 ){\makebox(0,0){\sfbtenpt{$N$}}}
  \put(+0.95  ,+0.225){\makebox(0,0){\sfbtenpt{$t$}}}
\end{picture}}
}
\newcommand{\LabelR}[1]{%
\makebox[0in]{
\setlength{\unitlength}{\Width}
\begin{picture}(0,0)
  \put(+0.09  ,+0.725){\makebox(0,0){\sfbtenpt{\shortstack[c]{#1}}}}
  \put(+0.3   ,+0.05 ){\makebox(0,0){\sfbtenpt{$N\!:2N$}}}
  \put(+0.95  ,+0.225){\makebox(0,0){\sfbtenpt{$t$}}}
\end{picture}}
}
\begin{center}
\begin{tabular}{||c|c||} \hline\hline
\LabelQ{$\lonenorm{E^{(N)}}$}\PSFIG{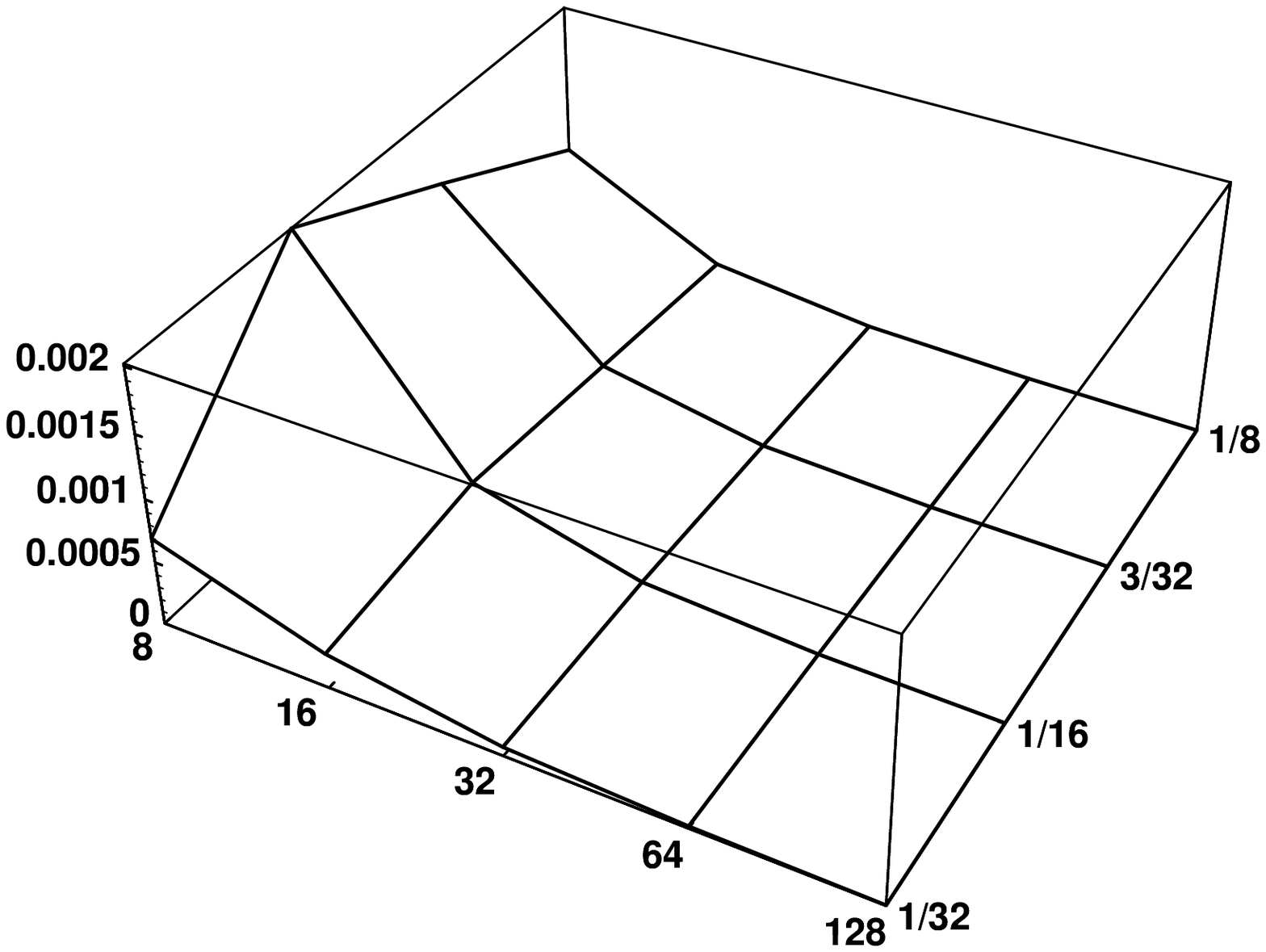} &
\LabelR{$\frac{\lonenorm{E^{(N)}}}{\lonenorm{E^{(2N)}}}$}\PSFIG{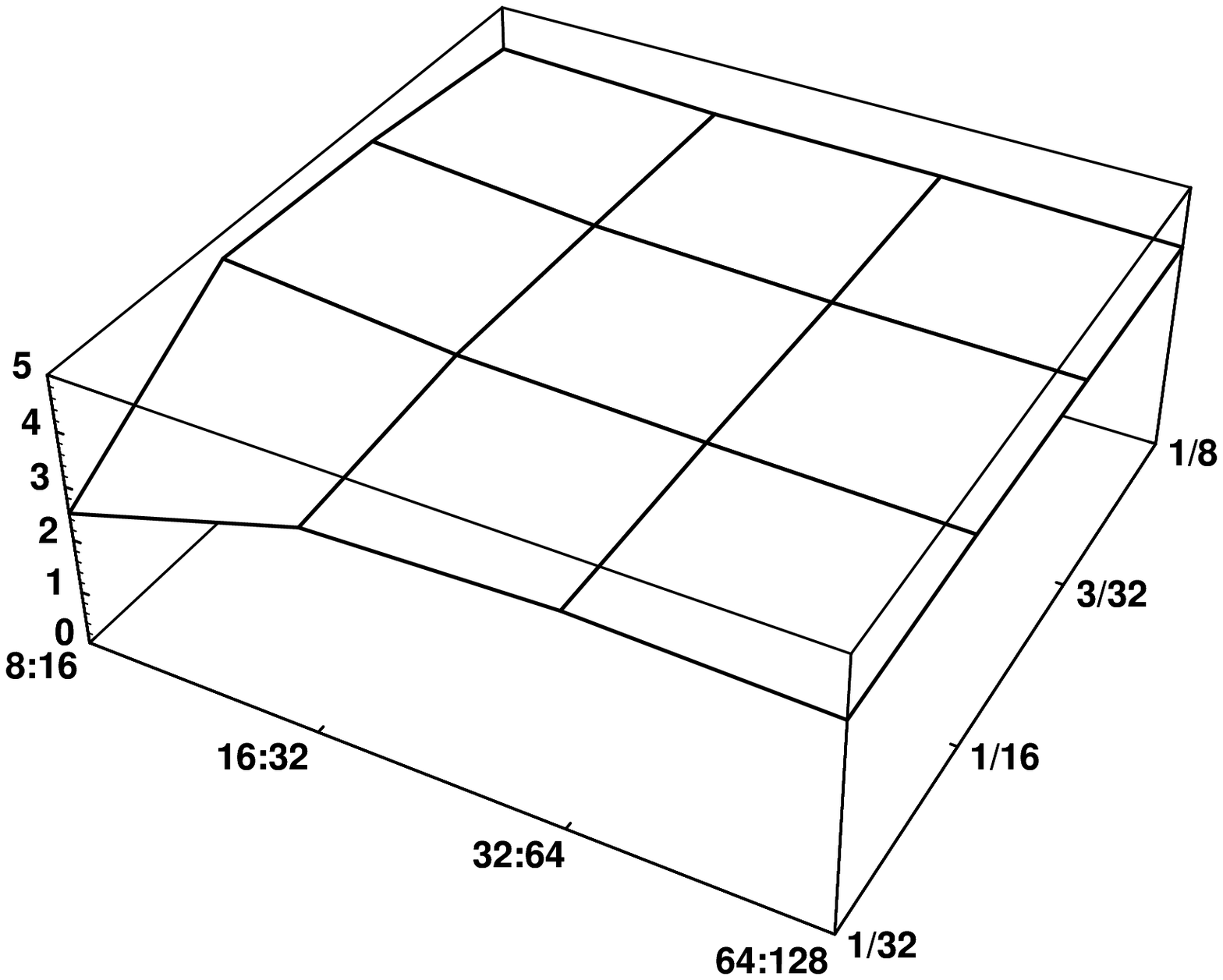}
\\
(a) & (b) \\ \hline\hline
\end{tabular}
\end{center}
\caption{%
LB2:~~Norm comparisons for $A=1/12$ and $B = 3/4$.
(a) Errors; (b) error ratios.
}
\label{fig9}
\end{figure}


\clearpage
\section{Example - Burgers' Equation}

  Boghosian and Levermore \cite{BL87} introduced a lattice Boltzmann
method for solving the one-dimensional viscous Burgers equation
\begin{equation}
  \label{burger}
  \rho_t + \rho \rho_x = \nu \rho_{xx} \,.
\end{equation}
The lattice in this case is one-dimensional and periodic, and the
particle states are given in Figure \ref{states7}.  The collision
rules are listed in Figure \ref{fig5}, where the probability of an
advection to the right, i.e., in the direction of $v=1$, is
$a=\tfrac12 (1+\eps)$ and to the left in the direction of $v=-1$ is
$\overline{a}=1-a$.
\begin{figure}
\def\PICIN#1#2{{
        \thicklines
        \setlength{\unitlength}{5em}
        \begin{picture}(1,0.5)(0,-0.2)
                \put(0.5,-0.1){\line(0,1){0.2}}
                \put(0.0,0.0){#1(+1,0){0.5}}
                \put(1.0,0.0){#2(-1,0){0.5}}
        \end{picture}
}}
\begin{center}
\begin{tabular}{||c|c||} \hline\hline
\multicolumn{2}{||c||}{Particle State $p$}      \\ \hline
   1                   &   -1                   \\ \hline\hline
\PICIN{\line}{\vector} & \PICIN{\vector}{\line} \\ \hline\hline
\end{tabular}
\end{center}
\caption{Particle States - 1-D Burgers' Equation.}
\label{states7}
\end{figure}
\begin{figure}
\def\PICINN#1#2{{
        \thicklines
        \setlength{\unitlength}{5em}
        \begin{picture}(1,0.5)(0,-0.2)
                \put(0.5,-0.1){\line(0,1){0.2}}
                \put(0.0,0.0){#1(+1,0){0.5}}
                \put(1.0,0.0){#2(-1,0){0.5}}
        \end{picture}
}}
\def\PICOUT#1#2{{
        \thicklines
        \setlength{\unitlength}{5em}
        \begin{picture}(1,0.5)(0,-0.2)
                \put(0.5,-0.1){\line(0,1){0.2}}
                \put(0.5,0.0){#1(-1,0){0.5}}
                \put(0.5,0.0){#2(+1,0){0.5}}
        \end{picture}
}}
\def\PROB#1{{
        \setlength{\unitlength}{5em}
        \begin{picture}(0,0)
                \put(0,0){\makebox(0,0){#1}}
        \end{picture}
}}
\def\PROB#1{\raisebox{0.75em}{#1}}
\def\Entry#1#2#3{
        {#1} & {#2} & \PROB{#3}
}
\begin{center}
\begin{tabular}{||c||c|c||} \hline\hline
Pre-Collision &
\multicolumn{2}{c||}{Post-Collision}
\\ \hline
$n$ & $n'$ & $\SS(n,n')$
\\ \hline\hline
\Entry  {\PICINN{\line}{\line}}
        {\PICOUT{\line}{\line}}{$1$}
        \\ \hline
\Entry  {\PICINN{\vector}{\line}}
        {\PICOUT{\vector}{\line}}{$\displaystyle{a}=\tfrac12(1+\eps)$}
        \\ \hline
\Entry  {\PICINN{\vector}{\line}}
        {\PICOUT{\line}{\vector}}{$\displaystyle\overline{a}=\tfrac12(1-\eps)
$}
        \\ \hline
\Entry  {\PICINN{\line}{\vector}}
        {\PICOUT{\vector}{\line}}{$\displaystyle{a}=\tfrac12(1+\eps)$}
        \\ \hline
\Entry  {\PICINN{\line}{\vector}}
        {\PICOUT{\line}{\vector}}{$\displaystyle\overline{a}=\tfrac12(1-\eps)
$}
        \\ \hline
\Entry  {\PICINN{\vector}{\vector}}
        {\PICOUT{\vector}{\vector}}{$1$}
        \\ \hline\hline
\end{tabular}
\end{center}
\caption{Collision Rules --- Burgers' Equation.}
\label{fig5}
\end{figure}
Here, $0<\eps<1$ is given.  The generalized Boltzmann collision
operator is
\begin{eqnarray}
  \label{coll22}
  \CC(F)
  &=& \tfrac12 [ -F_1 + F_{-1} + \eps (F_1 + F_{-1} - 2F_1 F_{-1}) ]
      \left[ \begin{array}{cc} +1 \\ -1 \end{array} \right]
\\
  \nonumber
  &\equiv& \CC^{(0)}(F) + \frac{\eps}{\KA} \CC^{(1)}(F) \,,~ \KA > 0 \,.
\end{eqnarray}
Thus, if we assume $\eps=\KA\delta$, then
$$
  \CC(F) = \CC^{(0)}(F) + \delta \CC^{(1)}(F) \,.
$$
Clearly, $\CC^{(0)}$ is a Boltzmann collision operator in semidetailed
balance.

  We have at an equilibrium
\begin{equation}
  \CC^{(0)}(F)(1) = \CC^{(0)}(F)(-1) \,,
\end{equation}
so that
$$
  F_1 = F_{-1} \equiv u \,.
$$
The linearized collision operator of $\CC^{(0)}$ is
$$
  \LL = \tfrac12
        \left[ \begin{array}{cc} -1 & +1 \\
                                 +1 & -1 \end{array} \right] \,.
$$
The eigenvalues of $\LL$ are given by $(\lambda_0,\lambda_1)=(0,-1)$,
and the unnormalized eigenmatrix is
$$
  {\bf Q} = [\q_0, \q_1]
  = \left[ \begin{array}{cc} +1 & +1 \\
                             +1 & -1 \end{array} \right] \,.
$$
The pseudoinverse of $\LL$ is
$$
  \LL^+ = -\tfrac12\q_1\q_1^t
  = -\tfrac12 \left[ \begin{array}{cc} +1 & -1 \\
                                       -1 & +1 \end{array} \right] \,.
$$

We determine the Hilbert expansion (\ref{series}) as before.  In this
case,
$$
  h^{(0)} = u \q_0 \,,
$$
and for $j = 1,2,3,4$
$$
  h^{(j)} = \beta^{(j)} \q_0 + c^{(j)} {\bf q}_1 \,.
$$
The $c^{(j)}$ are given in Appendix B.  Letting
$$
  A(u) = u(1-u) \,, \qquad
  c = \frac{\KA L}{T} \,,\qquad
  \mu= \frac{L^2}{2T} \,,
$$
we have
\begin{eqnarray}
  \label{eqqq}
  \del_t u + c \del_x A(u) & = & \mu \del_{xx} u \,,
\\
  \nonumber
  \beta^{(1)} & = & 0 \,,
\\
  \nonumber
  \del_t \beta^{(2)} + 6 c \del_x (A'(u) \beta^{(2)})
  & = & \mu \del_{xx} \beta^{(2)} - {\cal F} \,,
\\
  \nonumber
  \beta^{(3)} & = & 0 \,,
\end{eqnarray}
where ${\cal F}$ is a smooth function of $u$ and its derivatives.

Consider the grid function
$$
  H^{(4)}(t,x,\cdot) = \sum_{k=0}^4h^{(k)}(t,x,\cdot) \delta^k.~~
$$
where the $h^{(k)}$ are defined from the Hilbert expansion.
Then
$$
  \AA H^{(4)} - H^{(4)} - \CC(H^{(4)}) =  {\bf T}(H^{(4)})
  + {\rm higher~order~terms}
$$
where
$$
  {\bf T}(H^{(4)}) = \sum_{j=0}^4 {\bf T}^{(j)} \delta^j.
$$
It follows that ${\bf T}^{(j)} =0$ $j=0,1,2,3$.  Moreover,
${\bf T}^{(4)} = 0$ so that
$$
  \AA H^{(4)} - H^{(4)} - \CC(H^{(4)}) = {\cal O}(\delta^5)
$$
and the method is consistent.  Stability follows from the next
theorem.

\begin{theorem}
The set $\EE = [0,1] \times [0,1]$ is a domain of monotonicity
for (\ref{coll22}).
\end{theorem}

\begin{proof}
First note that
\[
\begin{array}{lcl}
  \JJ[\BB]_{1,1} & = & \tfrac12(1+\eps) - \eps F_1 \\
  \JJ[\BB]_{1,2} & = & \tfrac12(1+\eps) - \eps F_{-1} \\
  \JJ[\BB]_{2,1} & = & \tfrac12(1 - \eps) + \eps F_1 \\
  \JJ[\BB]_{2,2} & = & \tfrac12(1 - \eps) + \eps F_{-1}.
\end{array}
\]
Since $0<\eps<1$, we see that $\JJ[\BB] \geq 0$ on $\EE$.

  The extreme points of $\EE$ are
$$
  \MM_+ = \left[ \begin{array}{c} 0 \\ 0 \end{array} \right] \,, \qquad
  \MM_- = \left[ \begin{array}{c} 1 \\ 1 \end{array} \right] \,.
$$
Clearly, $\CC(\MM_+) = \CC(\MM_-) = 0$ and the
theorem is proved.
\end{proof}

\subsubsection*{Numerical Verification}

  We compute the solutions to the Burgers equation (\ref{eqqq})
for $x \in [0,1],~t > 0,~c=1,~\nu = 2^{-8},$ periodic boundary
conditions, and initial condition
$$
  u(0,x) = \tfrac{1}{2} \cos(2\pi x) + \tfrac{1}{2} \,.
$$
The finite difference solution $v_{{\bf i}}^m$ of (\ref{eqqq}) is
computed on a grid of size $N=32768$ by solving Burgers' equation
(\ref{burger}) with the conservative monotone difference scheme
$$
  \rho_i^{m+1}
  = \rho_i^m - \frac{\Delta t}{4 \Delta x}
               \left[ (\rho_{i+1}^m)^2 - (\rho_{i-1}^m)^2 \right]
             + \frac{\nu \Delta t}{(\Delta x)^2}
               \left[ \rho_{i+1}^m - 2 \rho_i^m + \rho_{i-1}^m \right] \,
$$
and then applying the transformation $\rho = 1-2v$.  The difference
method is first order accurate in time and second order accurate in
space and has the stability criteria $\Delta t\leq(\Delta x)^2/(2\nu)$.
Figure \ref{fig7} exhibits the initial condition and the solution at
time $t=\tfrac{1}{4}$.

  The lattice Boltzmann solutions $((u^N)_{{\bf i}}^m$ are computed on
grids of size $N=128$, $160$, $192$, $256$.  Here,
$u_{{\bf i}}^m =\tfrac12 \<F\>$.  Figure \ref{fig6} exhibits a
comparison of the finite difference- and lattice Boltzmann-computed
solutions, $V(t,x)$ and $U(t,x)$, respectively, at $t = \tfrac{1}{4}$.
This comparison indicates the importance of the underlying assumption
that $\eps={\cal O}((\Delta x)^2)$, which weakens as the grid is
coarsened.

  Table \ref{tab6} lists the $\ell_1$-norm of the error at time
$t=\tfrac{1}{4}$.  Varying in the table is the grid size for
the lattice Boltzmann method.  Focusing on the ratio
$\|E^{(N)}\|_{\ell_1} / \|E^{(2N)}\|_{\ell_1}$, the $\ell_1$-norm
results in the table strongly support the theoretical $O(\delta^2)$
convergence of the lattice Boltzmann method.  This point is also
illustrated in Figure \ref{fig10}.  The trend towards the value $4$ of
this quotient is apparent.
\begin{figure}
\def\Width{0.45\textwidth}
\def\Psfig#1#2#3{%
%
{\psfig{figure=psfigs/BL87/#1,rwidth=#2,height=#2,rheight=#3,bbllx=42.6483bp,bblly=36.4083bp,bburx=583.926bp,bbury=577.686bp,clip=}}
}
\begin{center}
\begin{tabular}{||c|c||} \hline\hline
%
\Psfig{}{\Width}{\Width} &
\Psfig{}{\Width}{\Width} \\
(a) & (b) \\ \hline\hline
\end{tabular}
\end{center}
\caption{Burgers' Equation: (a) $t=0$; (b) $t= \tfrac{1}{4}$.}
\label{fig7}
\end{figure}
\begin{figure}
\def\Width{0.45\textwidth}
\def\WidthR{0.45\textwidth}
\def\Psfig#1#2#3{%
%
{\psfig{figure=psfigs/BL87/#1,rwidth=#2,height=#2,rheight=#3,bbllx=42.6483bp,bblly=36.4083bp,bburx=605.26125bp,bbury=577.686bp,clip=}}
}
\begin{center}
\begin{tabular}{||c||} \hline\hline
\Psfig{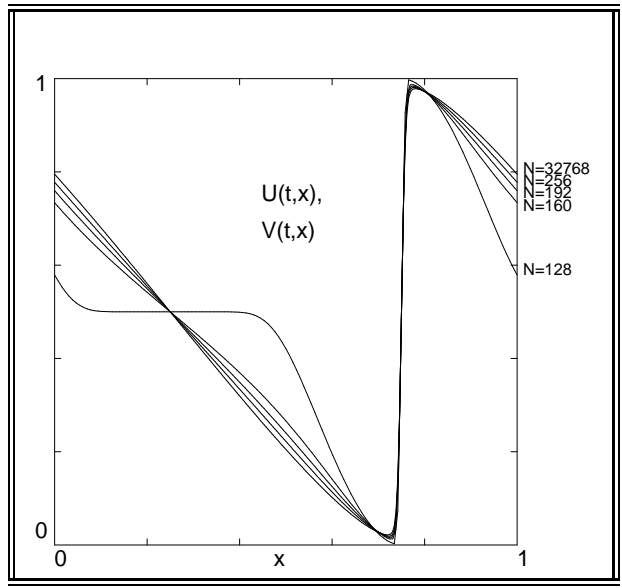}{\Width}{\WidthR}
\\*[-\baselineskip] \hline\hline
\end{tabular}
\end{center}
\caption{%
Burgers' Equation: Finite Difference and Lattice Boltzmann Solutions
at $t=\tfrac{1}{4}$.  The Finite Difference Solution $V(t,x)$ is the case in
which $N=32768\,$.
}
\label{fig6}
\end{figure}
%
%
\begin{table}
\def\FixUpRatio#1{\raisebox{1ex}[4ex]{#1}}
\def\Times{\!\!\times\!\!}
\def\norm#1{\left\|{#1}\right\|}
\def\Norm#1#2{\norm{#1}_{#2}}
\def\lonenorm#1{\Norm{#1}{\ell_1}}
\def\linfnorm#1{\Norm{#1}{\ell_\infty}}
\caption{Burgers' Comparison:~~Norm comparisons.}
\label{tab6}
\begin{center}
\renewcommand{\arraystretch}{1.25}
\begin{tabular}{||r||l|l|l|l||} \hline\hline
\mbox{} &
\multicolumn{4}{|c||}{\FixUpRatio{$\lonenorm{E^{(N)}}$}} \\ \cline{2-5}
\multicolumn{1}{||c||}{$N$} &
      $t=1/32$   & $t=1/16$   & $t=3/32$   & $t=1/8$    \\ \hline\hline
\long\def\Comment#1{}
\Comment{
0.0042538009583950043
0.00098613998852670193
0.00024224417575169355
0.000060288264648988843
0.000015040751350170467
0.0000037545605664490722
0.0000009295513905271946

0.0077094845473766327
0.0017746977973729372
0.00043512461706995964
0.00010823395859915763
0.000027015979867428541
0.0000067412784119369462
0.0000016782269085524604

0.0091696493327617645
0.0021021380089223385
0.00051491847261786461
0.00012802825949620456
0.000031939067412167788
0.0000079674509834148921
0.0000019765393517445773

0.010419003665447235
0.002348750364035368
0.00057307560928165913
0.00014235093840397894
0.000035512930480763316
0.0000088650740508455783
0.0000022066851670388132}
  256 & $4.254\Times10^{-3}$ & $7.709\Times10^{-3}$ &
        $9.170\Times10^{-3}$ & $1.042\Times10^{-2}$ \\ \hline
  512 & $9.861\Times10^{-4}$ & $1.775\Times10^{-3}$ &
        $2.102\Times10^{-3}$ & $2.349\Times10^{-3}$ \\ \hline
 1024 & $2.422\Times10^{-4}$ & $4.351\Times10^{-4}$ &
        $5.149\Times10^{-4}$ & $5.731\Times10^{-4}$ \\ \hline
 2048 & $6.029\Times10^{-5}$ & $1.082\Times10^{-4}$ &
        $1.280\Times10^{-4}$ & $1.424\Times10^{-4}$ \\ \hline
 4096 & $1.504\Times10^{-5}$ & $2.702\Times10^{-5}$ &
        $3.194\Times10^{-4}$ & $3.551\Times10^{-5}$ \\ \hline
 8192 & $3.755\Times10^{-6}$ & $6.741\Times10^{-6}$ &
        $7.967\Times10^{-5}$ & $8.865\Times10^{-6}$ \\ \hline
16384 & $9.296\Times10^{-7}$ & $1.678\Times10^{-6}$ &
        $1.977\Times10^{-5}$ & $2.207\Times10^{-6}$ \\ \hline
\multicolumn{5}{c}{\mbox{}\vspace{-2ex}} \\ \hline
\multicolumn{1}{||c||}{$N$} &
\multicolumn{4}{|c||}
               {\FixUpRatio{$\lonenorm{E^{(N)}}/\lonenorm{E^{(2N)}}$}}
\\ \hline
  256 & 4.314 & 4.344 & 4.362 & 4.436 \\ \hline
  512 & 4.071 & 4.079 & 4.082 & 4.098 \\ \hline
 1024 & 4.018 & 4.020 & 4.022 & 4.026 \\ \hline
 2048 & 4.008 & 4.006 & 4.009 & 4.008 \\ \hline
 4096 & 4.006 & 4.008 & 4.009 & 4.006 \\ \hline
 8192 & 4.039 & 4.017 & 4.031 & 4.017 \\ \hline\hline
\end{tabular}
\end{center}
\end{table}
\begin{figure}
\def\norm#1{\left\|{#1}\right\|}
\def\Norm#1#2{\norm{#1}_{#2}}
\def\lonenorm#1{\Norm{#1}{\ell_1}}
\def\linfnorm#1{\Norm{#1}{\ell_\infty}}
\def\sfbtenpt{\sf}
\setlength{\Width}{0.45\textwidth}
\newcommand{\PSFIG}[1]{%
%
{\psfig{figure=psfigs/#1,bbllx=72bp,bblly=72bp,bburx=540bp,bbury=605bp,width=\Width,rheight=0.85\Width,clip=}}%
}
\newcommand{\LabelQ}[1]{%
\makebox[0in]{
\setlength{\unitlength}{\Width}
\begin{picture}(0,0)
  \put(+0.09  ,+0.7  ){\makebox(0,0){\sfbtenpt{\shortstack[c]{#1}}}}
  \put(+0.3   ,+0.05 ){\makebox(0,0){\sfbtenpt{$N$}}}
  \put(+0.95  ,+0.225){\makebox(0,0){\sfbtenpt{$t$}}}
\end{picture}}
}
\newcommand{\LabelR}[1]{%
\makebox[0in]{
\setlength{\unitlength}{\Width}
\begin{picture}(0,0)
  \put(+0.09  ,+0.7  ){\makebox(0,0){\sfbtenpt{\shortstack[c]{#1}}}}
  \put(+0.3   ,+0.05  ){\makebox(0,0){\sfbtenpt{$N\!:2N$}}}
  \put(+0.95  ,+0.225){\makebox(0,0){\sfbtenpt{$t$}}}
\end{picture}}
}
\begin{center}
\begin{tabular}{||c|c||} \hline\hline
\LabelQ{$\lonenorm{E^{(N)}}$}\PSFIG{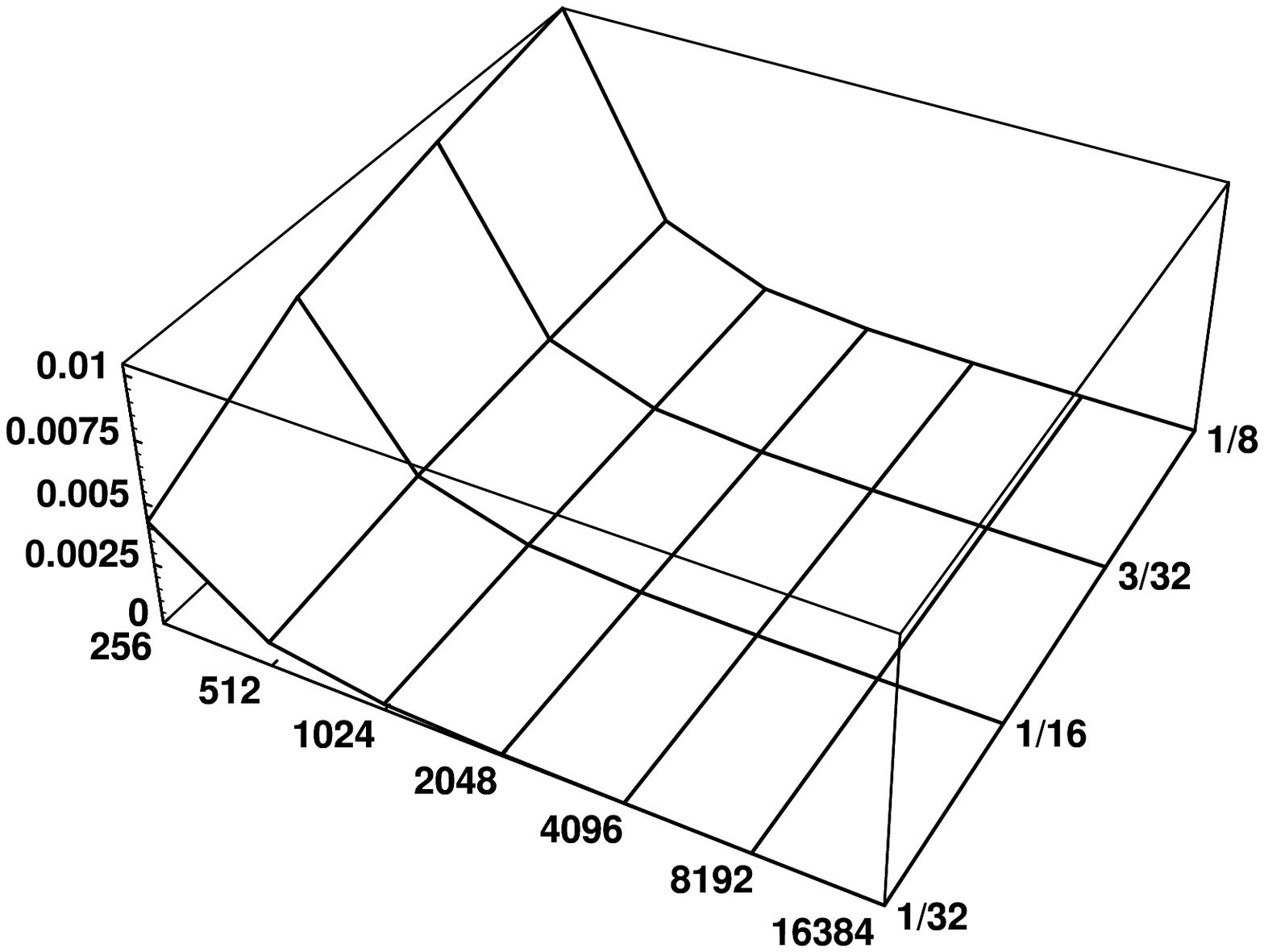} &
\LabelR{$\frac{\lonenorm{E^{(N)}}}{\lonenorm{E^{(2N)}}}$}
       \PSFIG{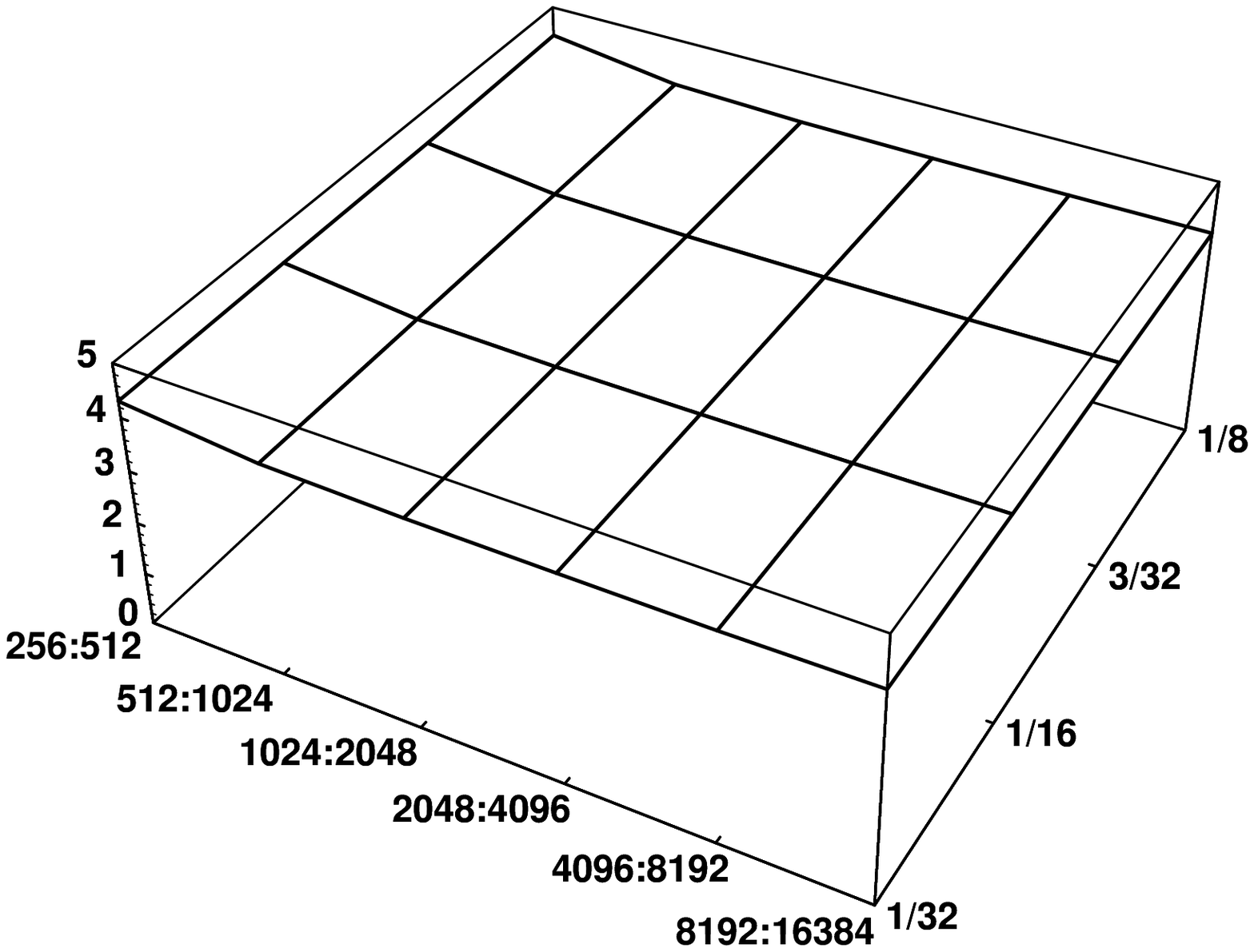} \\*[-1ex] (a) & (b)
\\ \hline\hline
\end{tabular}
\end{center}
\caption{%
Burgers' Equation:~~Norm comparisons. (a) Errors; (b) error ratios.
}
\label{fig10}
\end{figure}


\clearpage
\section{Conclusion}

  We defined a lattice Boltzmann method as an approximation to an
ensembled lattice gas method. The concept of semidetailed balance for
a lattice Boltzmann collision operator was defined and analyzed. This
property allowed us to prove an $H$-theorem which characterized the
equilibria of a Boltzmann collision operator.  An asymptotic Hilbert
expansion was constructed about an equilibrium solution of a diffusive
collision operator.  Convergence of a lattice Boltzmann method was
established by analyzing the behavior of a truncated Hilbert expansion
as the perturbation parameter approaches zero.  Stability, consistency
and convergence of a lattice Boltzmann method were defined.  Stability
and consistency were shown to imply convergence.  Monotone Boltzmann
collision operators were also defined and shown to imply stability.

  Three example lattice Boltzmann methods were analyzed and shown to
be consistent and stable. These properties allowed us to show that the
solutions converged; one to the solution of Burgers' equation and the
others respectively to the solutions of two nonlinear diffusion
equations.  Numerical results were presented that verified the
convergence of each of these methods.



\clearpage

\appendix
\ifdopreprint\else
\addtocontents{toc}{\protect\contentsline{section}{Appendices:}{}}
\fi

\section{Details of LB1 and LB2 Examples}

For LB1  we have
$$
  h^{(j)} = \beta^{(j)}{\bf q}_0 +  \sum_{k=1}^3 \nu^{-1} c_k^{(j)} \q_k \,,
$$
where $\nu = -4u(1-u)$ and (using the standard notation
$\GRAD = (\del_x,\del_y)$ and $\overline{\GRAD} = (\del_x,-\del_y)$)
\begin{enumerate}
\item ($j=0$);
$$
c_1^{(0)} = c_2^{(0)} = c_3^{(0)} = 0 \,.
$$
\item ($j=1$);
\begin{eqnarray*}
c_1^{(1)} &=& L  u_x \,, \\
c_2^{(1)} &=& L u_y \,, \\
c_3^{(1)} &=& 0 \,.
\end{eqnarray*}
\item ($j=2$);
\begin{eqnarray*}
c_1^{(2)} &=&  c_2^{(2)} = 0 \,, \\
c_3^{(2)} &=&
    \tfrac12 L^2  \left[ \tfrac12 D'(u) \GRAD u \cdot \overline{\GRAD} u
		 -  \overline{\GRAD} \cdot D(u) \GRAD u \right] .
\end{eqnarray*}
\item ($j=3$);
\begin{eqnarray*}
c_1^{(3)} &=& \tfrac16 L^3u_{xxx} + LTu_{xt}
+ \left [ \tfrac12 L^3 \partial_{xx}
+ LT \partial_t \right ] (\nu^{-1} u_x) \\
	&+&\tfrac12 L^3 \del_x
		\left[ (2\nu)^{-1} D'(u) \GRAD u \cdot \overline{\GRAD} u
		- \nu^{-1}\overline{\GRAD} \cdot D(u) \GRAD u \right] \\
	&+& L\beta^{(2)}_x + 4L\nu^{-3} (Lu_y)^2  \\
	&+& \tfrac14 L^3 D'(u)  u_x
		\left [  D'(u) \GRAD u \cdot \overline{\GRAD} u
		-2 \overline{\GRAD} \cdot D(u) \GRAD u \right ] ; \\
c_2^{(3)} &=& \tfrac16 L^3 u_{yyy} + LTu_{yt}
		+ \left ( \tfrac12 L^3 \del_{yy} + LT \del_t \right )
			(\nu^{-1} u_y) \\
	&-& \tfrac12 L^3 \del_y
		\left[ (2\nu)^{-1} D'(u) \GRAD u \cdot \overline{\GRAD} u
		  - \nu^{-1} \overline{\GRAD}\cdot D(u) \GRAD u \right] \\
	&+& L\beta^{(2)} + 4L^3 \nu^{-3}(u_x)^2 \\
	&-&  \tfrac14 L^3 D'(u) u_y
		\left[ D'(u) \GRAD u \cdot \overline{\GRAD} u
		-2 \overline{\GRAD} \cdot D(u) \GRAD u \right] ; \\
c_3^{(3)} &=& 0 \,.
\end{eqnarray*}
\end{enumerate}

\noindent For LB2 we have
$$
h^{(j)} = \beta^{(j)} {\bf q}_0
	+ \sum_{k=1}^3 \lambda_k^{-1} c_k^{(j)} {\bf q}_k~,
$$
where $\lambda_1 = \lambda_2 = \nu = -2(1-u^2),~\lambda_3 = -4u(1-u)$ and
\begin{enumerate}
\item ($j=0$);
$$
c_1^{(0)} = c_2^{(0)} = c_3^{(0)} = 0~.
$$
\item ($j=1$);
\begin{eqnarray*}
c_1^{(1)} &=& Lu_x~, \\
c_2^{(1)} &=& Lu_y~, \\
c_3^{(1)} &=& 0~.
\end{eqnarray*}
\item ($j=2$);
\begin{eqnarray*}
c_1^{(2)} &=& c_2^{(2)} = 0~, \\
c_3^{(2)} &=& \tfrac12 L^2
	\left [ (2u-1) \nu^{-2} \nabla u \cdot \overline{\nabla} u
		- \overline{\nabla} \cdot D(u) \nabla u \right ]~.
\end{eqnarray*}
\item ($j=2$);
\begin{eqnarray*}
c_1^{(3)} &=& \tfrac16 L^3 u_{xxx} + LT u_{xt} + \left [ \tfrac12 \partial_{xx}
	+LT\partial_t \right ] (\nu^{-1} u_x) \\
&+& \tfrac12 L^3 \partial_x \nu^{-1} \partial_x u
	 + 2 \nu^{-3} L^3 u_x (u_y)^2 + L\beta_x^{(2)} \\
&-& LD'(u) \nu \beta^{(2)} u_x
- L^2 u_x \overline{\nabla} \cdot D(u) \nabla u
+2\nu^{-1}L^2(2u-1) u_x \overline{\nabla}\cdot \nabla u \\
&+& L^3 \partial_x
	\left [ \tfrac12 \lambda_3^{-1} \overline{\nabla} \cdot D(u) \nabla u
	+ \nu^{-2} \lambda_3^{-1} (2u-1) \overline{\nabla} \cdot \nabla u
	\right ]~, \\
c_2^{(3)} &=& \tfrac16 L^3 u_{yyy} +LTu_{yt}
+ \left [ \tfrac12 L^3 \partial_{yy}
+LT \partial_t \right ] (\nu^{-1} u_x ) \\
&+& \tfrac12 L^3 \partial_y \nu^{-1} \partial_y u
+ 2 \nu^{-3} L^3 u_x^2 + L \beta_y^{(2)} \\
&-& LD'(u) \nu^{-1} \beta^{(2)} u_y \\
&+& L^2 u_y \left [ \overline{\nabla} \cdot D(u) \nabla u + 2 \nu^{-1} L^2
(2u-1)
u_y \overline{\nabla} \cdot \nabla u \right ] \\
&-& L^3 \partial_y \left [
	  \tfrac12 \lambda_3^{-1} \overline{\nabla}\cdot D(u) \nabla u
	+ \nu^{-2} \lambda_3^{-1} (2u-1) \overline{\nabla} \cdot \nabla u
	\right ]~.
\end{eqnarray*}
\end{enumerate}


\clearpage

\section{Details of the Burgers Equation Example}

For $h^{(j)} = \beta^{(j)} {\bf q}_0 + c^{(j)} {\bf q}_1$ in Burgers' equation,
\begin{enumerate}

\item
$$
  c^{(1)} = \KA A(u) - Lu_x \,;
$$

\item
$$
  c^{(2)} = 0 \,;
$$

\item
\begin{eqnarray*}
  c^{(3)}
  & = & - \tfrac13 L^3 u_{xxx}
        + \KA \left( \tfrac12 L^2 \del_x + T \del_t \right) A'(u)
        + L \beta^{(2)} - 6 \KA A'(u) \beta^{(2)}
\\
  &   & + \KA \left[ (\beta^{(2)})^2 - (L u_x - \KA A(u))^2 \right] \,;
\end{eqnarray*}

\item
$$
  c^{(4)} = 0 \,.
$$

\end{enumerate}


\clearpage

\section{Further Details of the LB1 and LB2 Examples}

For
$$
{\bf T}^{(4)} = - \tilde{c}^{(4)}_3 {\bf q}_3
$$
in LB1 we have
\begin{eqnarray*}
\tilde{c}_3^{(4)} &=& \tfrac1{48} L^4(u_{xxxx} - u_{yyyy})
		+ \tfrac14 L^2T(u_{xxt}-u_{yyt})
		+ T \del_t c_3^{(2)} \\
	&+& \tfrac14 L^2 \left [ \overline{\GRAD} \cdot \GRAD \beta^{(2)}
		+     \GRAD^2 c_3^{(2)} \right]
		+ \tfrac12  L \left ( \del_x c_1^{(3)}
		- \del_y c_2^{(3)} \right ) \\
	&+& \tfrac1{12} L^4
		(\del_{xxx}, -\del_{yyy} ) \cdot \del_t (\nu^{-1} \GRAD u)
		+ \tfrac12 L^2T \overline{\GRAD} \cdot \del_t(\nu^{-1} \GRAD u)
	\\
	&-& \tfrac12 L^4\nu^{-3} D'(u) \left ( (u_x)^4 - (u_y)^4 \right ) \\
	&-& L^4\nu^{-3} \left[( (u_x)^2
		 + (u_y)^2) \overline{\GRAD} \cdot D(u) \nabla u
		 - 2 \nu \beta^{(2)} \left ( (u_x)^2 - (u_y)^2 \right )
		\right] \\
	 &-& \tfrac14 L^2 D'(u) \nu \beta^{(2)} \left[
		  D'(u) \GRAD u \cdot \overline{\GRAD} u
		 -2 \overline{\GRAD} \cdot D(u) \GRAD u
		\right] \\
	 &+& \tfrac12 L D'(u) \nu \left ( c_1^{(3)} u_x - c_2^{(3)}u_y \right )
\end{eqnarray*}
and in LB2 we have
\begin{eqnarray*}
\tilde{c}_3^{(4)} &=& \tfrac1{48} L^4(u_{xxxx} - u_{yyyy})
		+ \tfrac14 L^2T(u_{xxt}-u_{yyt})  + T \del_t c_3^{(2)} \\
	&+& \tfrac14 L^4   \GRAD^2  c_3^{(2)}
		+ \tfrac12  L \left ( \del_x c_1^{(3)}
		- \del_y c_2^{(3)} \right ) \\
	&+& \tfrac1{12} L^4 (\del_{xxx}, -\del_{yyy} ) \cdot \nu^{-1 } \GRAD u
		+ \tfrac12 L^2T \overline{\GRAD} \cdot \del_t(\nu^{-1} \GRAD u)
	\\
	&+& 2L(2u-1)\nu^{-1} (c_1^{(3)} u_x - c_2^{(3)}) -   L^2\nu^{-3}  [
		2L^2(2u-1)\nu^{-1}\lambda_3^{-1}
		  \left ( (u_x)^4 - (u_y)^4 \right ) \\
	&-& L^2 \nu \lambda_3^{-1}( (u_x)^2 + (u_y)^2)
		\overline{\GRAD} \cdot D(u) \nabla u
		 - 2 \nu \beta^{(2)} \left ( (u_x)^2 - (u_y)^2 \right )
		] \\
	&-& 2 L^2(2u-1)\lambda_3^{-1} \beta^{(2)}
		\left[
			2(2u-1)\nu^{-2} \GRAD u \cdot \overline{\GRAD} u
				- \overline{\GRAD} \cdot D(u) \GRAD u
		\right] .
\end{eqnarray*}


\clearpage
\addcontentsline{toc}{section}{References}

\begin{thebibliography}{99}

\bibitem{BGL91}
C. Bardos, F. Golse \& D. Levermore,
{\it Fluid Dynamic Limits of Discrete Velocity Kinetic Equations\/},
in ``Advances in Kinetic Theory and Continuum Mechanics'',
     R. Gatignol \& Soubbaramayer (eds.), Springer (1991), 57--71.

\bibitem{BPT88}
N. Bellomo, A. Palczewski \& G. Toscani,
{\it Mathematical Topics in Nonlinear Kinetic Theory\/},
World Scientific (1988).

\bibitem{BL87}
B. Boghosian \& C.D. Levermore,
{\it A cellular automaton for Burgers' equation\/},
Complex Systems 1(1) (1987), 17--30.

\bibitem{BL89a}
B. Boghosian \& C.D. Levermore,
{\it A Deterministic Cellular Automaton with Diffusive Behavior\/},
in ``Discrete Kinetic Theory, Lattice Gas Dynamics
     and Foundations of Hydrodynamics'',
     R. Monaco (ed.), World Scientific (1989).

\bibitem{BL89b}
B. Boghosian \& C.D. Levermore,
{\it Deterministic Cellular Automata with Diffusive Behavior\/},
in ``Cellular Automata and Modeling of Complex Physical Systems'',
     P. Manneville, N. Boccara, G.Y. Vichniac \& R. Bidaux (eds.),
     Proceedings in Physics 46,
     Springer-Verlag (1989), 118--129.

\bibitem{CDDEFT90}
S. Chen, K. Diemer, G. Doolen, K. Eggert \& B. Travis,
{\it Lattice gas automata for flow through porous media\/},
Physica D 47 (1991) 72--74.

\bibitem{CHL91}
R. Cornubert, D. d'Humieres \& C.D. Levermore,
{\it A Knudsen layer theory for lattice gases\/},
Physica D 47 (1991) 241--259.

\bibitem{Doolen91}
G.D. Doolen (ed.),
``Lattice Gas Methods for PDE's: Theory, Application, and Hardware'',
Physica D 47(1--2) (1991).
A comprehensive list of references appears on pp.\ 299--337.

\bibitem{Elton90}
A.B.H. Elton,
{\it A Numerical Theory of Lattice Gas and Lattice Boltzmann Methods
     in the Computation of Solutions to Nonlinear Advective-Diffusive
     Systems\/},
Ph.D. Thesis, University of California, Davis, CA, Sept. 1990,
Lawrence Livermore National Laboratory Report \#UCRL--LR--105090.

\bibitem{ELR90}
B.H. Elton, C.D. Levermore, \& G. Rodrigue,
{\it Lattice Boltzmann methods for Some 2-D Nonlinear Diffusion
     Equations: Computational Results\/},
in ``Asymptotic Analysis and Numerical Solution
     of Partial Differential Equations'',
     H. Kaper (ed.),
     Lecture Notes in Pure and Applied Mathematics 130,
     Marcel Dekker Inc. (1990).

\bibitem{FHP86}
U. Frisch, B. Hasslacher, \& Y. Pomeau,
{\it Lattice gas automata for the Navier-Stokes equation\/},
Phys. Rev. Lett. 56 (1986) 1505--1508.

\bibitem{FHHLPR87}
U. Frisch, D. d'Humieres, B. Hasslacher, P. Lallemand, Y. Pomeau,
\& J.P. Rivet,
{\it Lattice gas hydrodynamics in two and three dimensions\/},
Complex Systems 1(4) (1987) 599--707.

\bibitem{HPP76}
J. Hardy, O. de Pazzis \& Y. Pomeau,
{\it Molecular dynamics of a classical gas:
     Transport properties and time correlation functions\/},
Phys. Rev. A 13(5) (1976) 1949--1961.

\bibitem{HP72}
J. Hardy \& Y. Pomeau,
{\it Thermodynamics and hydrodynamics for a modeled fluid\/},
J. Math. Phys. 13(7) (1972) 1042--1051.

\bibitem{HPP73}
J. Hardy, Y. Pomeau \& O. de Pazzis,
{\it Time evolution of a two-dimensional model system I:
     Invariant states and time correlation functions\/},
J. Math. Phys. 14(12) (1973) 1746--1759.

\bibitem{Levermore92}
D. Levermore,
{\it Fluid Dynamical Limits of Discrete Kinetic Theories\/},
in ``Macroscopic Simulations of Complex Phenomena'',
     M. Mareschal \& B. Holian (eds.),
NATO ASI Series B, Plenum (1992), 173--185.

\bibitem{LCCDLR91}
L. Luo, H. Chen, S. Chen, G. Doolen, Y. Lee, \& H. Rose,
{\it Generalized hydrodynamic transport in lattice-gas automata\/},
Phys. Rev. A 43(12) (1991), 7097--7100.

\bibitem{MZ88}
G. McNamara \& G. Zanetti,
{\it Using the lattice Boltzmann equation
     to simulate lattice gas automata\/},
Phys. Rev. Lett. 61(20) (1988), 2332--2335.

\bibitem{MD87}
D. Montgomery \& G. Doolen,
{\it Two cellular automata for plasma computations\/},
Complex Systems 1(4) (1987) 831--838.

\bibitem{Muir87}
F. Muir, private communication (1987).

\bibitem{RM67}
R.D. Richtmyer \& K.W. Morton,
{\it Difference Methods for Initial-Value Problems\/},
Interscience Tracts in Pure and Applied Mathematics 4,
Interscience Publishers (John Wiley \& Sons), 2nd ed. (1967).

\bibitem{von.Neumann!cellular.automata}
J. von Neumann,
{\it Theory of Self-Reproducing Automata\/},
Univ. of  Ill. Press (Urbana), 1966.
See also A.W. Burkes (ed.), {\it Essays on Cellular Automata\/},
Univ. of Ill. Press (Urbana) (1970), p. 367.

\bibitem{Sod88}
G.A. Sod,
{\it Numerical Methods in Fluid Dynamics:
     Initial and Initial Boundary-Value Problems\/},
Cambridge University Press, Cambridge (1988).

\bibitem{SBH89}
S. Succi, R. Benzi \& F.$\,$J. Higuera,
{\it Lattice gas and Boltzmann simulations of homogeneous
     and inhomogeneous hydrodynamics\/},
in ``Discrete Kinetic Theory, Lattice Gas Dynamics
     and Foundations of Hydrodynamics'',
R. Monaco (ed.), World Scientific (1989) 329--342.

\bibitem{Wolfram86}
S. Wolfram,
{\it Cellular automaton fluids 1: Basic theory\/},
J. Stat. Phys. 45 (1986), 471--526.

\end{thebibliography}
\end{document}